\documentclass[referee]{raa}
\usepackage{graphicx,times}
\usepackage{natbib}
\usepackage{amssymb}

\usepackage{newtxtext}
\usepackage[T1]{fontenc}
\usepackage{ae,aecompl}
\usepackage{threeparttable}
\usepackage{ulem}
\usepackage{txfonts}
\usepackage{amssymb}	

\def\kms  {km~s$^{-1}$}

\bibpunct{(}{)}{;}{a}{}{,}

\usepackage{hyperref}
\usepackage{longtable}
\usepackage{booktabs}
\hypersetup{pdftitle = The title of my PDF, pdfauthor = My name, pdfsubject= The subject, pdfkeywords = keyword1 keyword2 keyword3} 
\hypersetup{colorlinks = true, linkcolor = green, anchorcolor = red, citecolor = blue, filecolor = red, pagecolor = red, urlcolor = blue}

\begin{document}

   \title{A catalogue of 74 new open clusters found in \emph{Gaia} Data-Release 2}

 \volnopage{ {\bf 2018} Vol.\ {\bf X} No. {\bf XX}, 000--000}
   \setcounter{page}{1}

   \author{Zhi-Hong He\inst{1,2}, Ye Xu\inst{1}, Chao-Jie Hao\inst{1,2}, Zhen-Yu Wu\inst{3,4}, Jing-Jing Li\inst{1}}

   \institute{Purple Mountain Observatory, CAS, No.10 Yuanhua Road, Qixia District, Nanjing 210023, China. 
\and University of Science and Technology of China, No.96, Jinzhai Road Baohe District, Hefei, Anhui, 230026, China\\
\and National Astronomical Observatories, CAS, 20A Datun Road, ChaoYang District, Beijing 100101, China\\
\and Key Laboratory of Optical Astronomy, National Astronomical Observatories, CAS, Beijing 100101, China\\
\vs \no
}

\abstract{
Based on astrometric data from \emph{Gaia} Data-Release 2 (DR2), we employ an unsupervised machine learning method to blindly search for open star clusters in the Milky Way within the Galactic latitude range of $|b| < 20^{\circ}$. In addition to 2,080 known clusters, 74 new open cluster candidates are found. In this work, we present the positions, apparent radii, parallaxes, proper motions and member stars of these candidates \footnotemark[1]\footnotetext[1]{
  \url{https://cdsarc.u-strasbg.fr/ftp/vizier.submit//new_OC/}}. Meanwhile, to obtain the physical parameters of each candidate cluster, stellar isochrones are fit to the photometric data. The results show that the apparent radii and the observed proper motion dispersions of these new candidates are consistent with those of open clusters previously identified in \emph{Gaia} DR2.
}
   \maketitle


\section{Introduction}\label{introduction}
 
An open cluster (hereafter OC) is a gravitationally bound stellar system composed of dozens to thousands of stars, which presents a loose structure and is irregularly shaped \citep{Zu03}. The member stars of an OC originated from the collapse of the same dense molecular environment \citep{McKee07}, and thus they have similar ages, kinematics and chemical compositions. The age of OCs can range from millions to billions of years \citep{Dias02, Kharchenko13,CG20}, making them ideal places to study the evolution of stars, and they are important tracers to study the structure and kinematics of the Milky Way \citep[e.g.][]{Friel95,Lada03,Carraro07,CG20,Pang20}.

It is estimated that the total number of OCs in the Galactic disk is of the order 10$^5$ \citep{Piskunov06}. The traditional method of identifying OCs is to use the positions and proper motions of groups of stars \citep[e.g.][]{Sanders71,Slovak77,Zhao90,Uribe94,Sanchez10}. In addition, photometric data can be used to make a colour-magnitude diagram (CMD) of a cluster to which stellar isochrones can be fit to obtain the physical parameters of the OC, such as its age, extinction and distance modulus. Before the \emph{Gaia} era, more than 3,000 OCs had been identified and catalogued using ground-based telescopes \citep[e.g.][]{Dias02, Kharchenko13}.

 \emph{Gaia} DR2 contains accurate parallaxes and proper motions for more than one billion stars \citep{Gaia18-Brown,Lindegren18}, where stars at different distances are distinguished via stellar parallaxes.  Based on \emph{Gaia} data, searches for OCs have yielded great results. At present, $\sim$1,500 known star clusters have been identified in \emph{Gaia} DR2 \citep[e.g.][]{CG18,Liu19}; in addition, $\sim$1,100 most probable new OC candidates have been recently published \citep[e.g.][]{CG18,CG19-0,Sim19,Liu19,Castro18,Castro19,Castro20,Ferreira19,Ferreira20,Hao20,Qin20}.

The continuous discovery of new OCs shows that not all existing star clusters have yet been found. In this paper, relying on the precise astrometric data of \emph{Gaia} DR2, we employ an unsupervised machine learning method to carry out blind searches for clusters within $|b| < 20^{\circ}$. After removing 2,080 previously reported clusters, we find 74 new OC candidates and obtain their physical parameters. 

 The remainder of this paper is organised as follows. In Section~\ref{data} we introduced the data. Section~\ref{method} presents our methods, including data preprocessing, the clustering algorithm and the isochrone fitting. Section~\ref{results} presents the results and provides discussions, and the conclusions are summarised in Section~\ref{sec:summary}.

\section{Data}\label{data}

The \emph{Gaia} satellite was launched by the European Space Agency in 2013, and its first data release was in 2016  \citep{Gaia16-Prusti}. DR2 \citep[][]{Gaia18-Brown} provided high-precision positions and $G$-band photometric data of approximately 1.7 billion sources, of which 1.4 billion sources have both $G_{\rm BP}$ and $G_{\rm RP}$ magnitudes, and 1.3 billion stars have measured parallaxes and proper motions. 

Previous studies reported that most known OCs are located near the Galactic plane $|b| < 20^{\circ}$ ~\citep[e.g.][]{Dias02, Kharchenko13}.
Hence, we applied three selection cuts to the \emph{Gaia} DR2 catalogue:

\begin{itemize}
\item[\textbullet] $|b| < 20^{\circ}$,
\item[\textbullet] $\varpi >$ 0.2~mas, parallax$\_$over$\_$error > 1,
\item[\textbullet] $G < 18$\,mag.
\end{itemize}

Thus, stars with $G < 18$\,mag and which possess a parallax uncertainty of $\sim$0.2~mas or better were added to our sample, as previously used by \citet{CG18,CG20,Liu19,Ferreira20}. The final sample contained 169,137,482 stars with astrometric parameters $(l, b, \varpi, \mu_{\alpha^*}, \mu_{\delta})$, where 166,136,789 had photometric data ($G$, $G_{\rm BP}$, $G_{\rm RP}$).

\section{Method}\label{method}

We used the stellar astrometric and photometric data derived in Section~\ref{data} to search for and identify new star clusters. The main method used is as follows: 

\begin{itemize}
\item[\textbullet] Preprocess the astrometric data $(l, b, \varpi, \mu_{\alpha^*},\mu_{\delta})$;
\item[\textbullet] Fit $k_{th}$NND (see Section~\ref{knnd}) to obtain the clustering parameters, and use the clustering algorithm to obtain cluster groups; and
\item[\textbullet] Fit stellar isochrones to the photometric data ($G$, $G_{\rm BP}$-$G_{\rm RP}$) of the cluster candidates.
\end{itemize}

\subsection{Astrometric data preparing}\label{preparing}

Following the same approach as \citet{Hao20}, the regions containing the data (as identified in Section~\ref{data}) were divided into rectangles of size $2^\circ \times  2^\circ   (|b| \leq 5^{\circ})$ and $3^\circ \times  3^\circ (|b| > 5^{\circ})$. Considering that the positions $(l, b)$ and proper motions $(\mu_{\alpha^*},\mu_{\delta})$ are relative parameters, i.e. the absolute values of distances and proper motions of stars with different parallaxes are variational, we used the projected positions, parallaxes and decomposed proper motions as input data for the clustering analysis as: 

\begin{equation}\label{eqn}
\left(d_{l^*}, d_b, \varpi, v_{\alpha^*}, v_{\delta}, \right) = (d\cdot\sin\theta_{l^*}, d\cdot\sin\theta_b, \varpi, d\cdot\mu_{\alpha^*}, d\cdot\mu_{\delta}),
\end{equation}

where the distance $d$ is taken to be the inverse of the parallax, 
 while $ \theta_{l^*}$ ($\theta_{l^*} = \theta_{l}\cdot\cos b$) and $\theta_{b}$ are the angular sizes of the star and the centre of the rectangle in $l$ and $b$, respectively. In addition, we calculated the median dispersion of ($d_{l^*},d_b, \varpi, v_{\alpha^*}, v_{\delta}$) for the 2,017 OCs catalogued by ~\citet{CG20} to standardize the above input data, so that the values in each dimension were multiples of the median dispersion of the corresponding parameter, and thus the weights of the input parameters in the process were equalized. Next, we used the DBSCAN algorithm to find clustered groups of stars.

\subsection{Clustering algorithm}\label{clustering}

\subsubsection{DBSCAN}\label{dbscan}

The unsupervised machine learning method DBSCAN used in this work was derived from \texttt{scikit-learn} \citep{Pedregosa11}. This algorithm was used to distinguish high-density groups from low-density regions in $n$-dimensional samples \citep{Ester96}, including extracting the core members and border members located in different groups, and rejecting any outliers not located in any group.  There are two parameters in the algorithm: the minimum number of data points $min\_samples$ and the radius  $\epsilon$, of which the latter was measured by a distance function as follows:

\begin{equation}\label{eq0}
D_{ij} = \sqrt{(x_{i1}-x_{j1})^2+(x_{i2}-x_{j2})^2+...+(x_{i5}-x_{j5})^2}
\end{equation}

where $D_{ij}$ is the distance between data points $x_{i}$ and $x_{j}$. 

A core is defined as a sub-sample of the data set where $min\_samples$ neighbours exist within a distance equal to $\epsilon$; that is, the core members are located in a dense area of vector space. Border members themselves do not form cores, but they may be located close to core members (i.e. $D_{ij}$ < $\epsilon$). In addition, outliers cannot constitute a core sample, nor can a candidate be composed of just border members, and they must be located at least $\epsilon$ from any core member. A larger $min\_samples$  or a smaller  $\epsilon$ means a higher density required to form a group.

\subsubsection{Clustering parameters: $min\_samples$ and  $\epsilon$}  \label{knnd}

The parameter $min\_samples$ mainly controls the tolerance to outliers in this algorithm. \citet[][]{Schubert17} found that $min\_samples$ has a weak effect on the clustering results. \citet{Sander98} suggested that $min\_samples$ can be determined as twice the value of the dimension of the data set; that is, $min\_samples$ =2 $\cdot$ $dim$, where $dim$ = 5 in this work.  We tried different values of $min\_samples$ around 10, and found that the detection efficiency was highest when using $min\_samples = 8$, that is, we could obtain more cores without obvious outliers.
 At the same time, this value was also within the range of the best values of $min\_samples$  used by \citet{Castro18, Hao20}. We present the results of $min\_samples$ = 8 in this article.

The parameter $\epsilon$ controls the radius of the local area of the points in the data set, and cannot be set to the default value~\citep{Schubert17, Castro18}. When this parameter is too small, most of the data will not be clustered; when it is too large, any adjacent  clusters or outliers will merge into one group. In previous studies, \citet{Ester96,Sander98} selected $\epsilon$ based on the $k_{th}$ nearest neighbour distance ($k_{th}$NND). On this basis, \citet[][]{Schubert17} obtained the corresponding $\epsilon$ value by observing the significant change of sorted $k_{th}$NND plot. In the algorithm of $k_{th}$NND, the reference point is not specified as a neighbor, while that is included in the DBSCAN for density estimation, so the $k$ value used in this work is $k$ = $min\_samples$ -1. 

\begin{figure*}
\begin{center}
	\includegraphics[width=1.\linewidth]{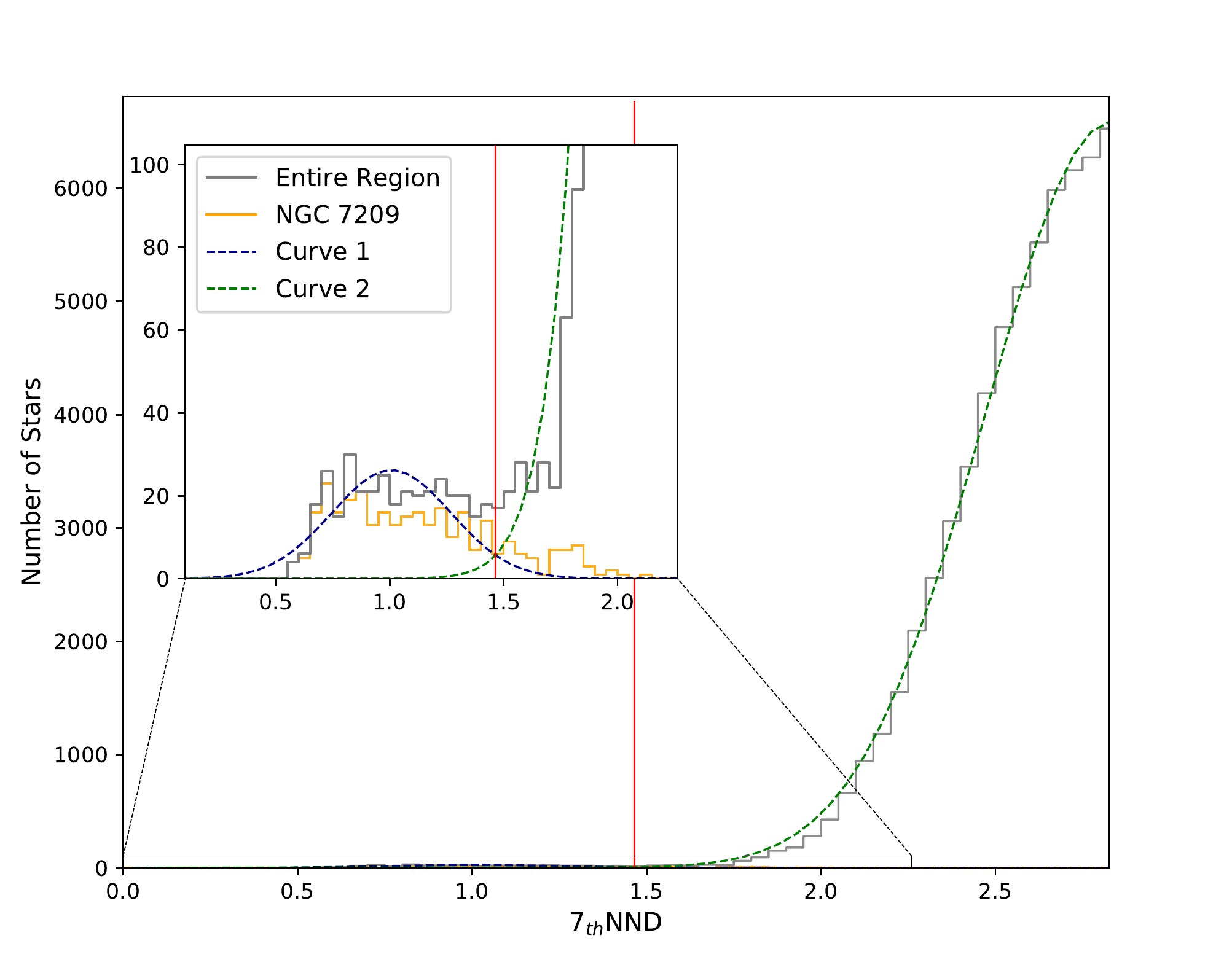}
	\caption{
	Example histogram of the $7_{th}$NND. The grey steps show the cumulative distribution of the star counts of $7_{th}$NND in the region around cluster NGC 7209. The orange steps shows the $7_{th}$NND histogram of member stars in NGC 7209, while the blue and green dashed lines show the Gaussian curves derived from Eq.~\ref{eqn1}. The red line corresponds to the chosen value of $\epsilon$  in this region, and the subplot shows a more visible zoomed-in region containing the bottom left end of the histogram.
	}
	\label{fig1}
\end{center}
\end{figure*}

In this work, the main steps to get $\epsilon$ are:

\begin{itemize}
\item Compute the $7_{th}$NND histogram via $ssklearn$ \footnotemark[1]\footnotetext[1]{
  \url{https://github.com/scikit-learn/scikit-learn}}. 

\item Fit the $7_{th}$NND using a bimodal Gaussian function. For gravitationally bound systems such as OCs, the intervals between the member stars in position and proper motion vectors are closer than those of field stars, so their $7_{th}$NNDs are significantly smaller than those of field stars \citep[][]{Castro18}. In the $7_{th}$NND histogram, the $7_{th}$NND of the cluster is distributed on the leftmost end (i.e. the minimum end, Fig.~\ref{fig1}) of the whole data set. Inspired by \citet[][]{Castro18}, in this work, we used a fitted $7_{th}$NND histogram to determine $\epsilon$ as: 

\begin{equation}\label{eqn1}
\mathbf{n}=a_1\cdot e^{\frac{-(\mathbf{d}_{k,nn}-\mu_1)^2}{2\sigma_1^2}}+a_2 \cdot e^{\frac{-(\mathbf{d}_{k,nn}-\mu_2)^2}{2\sigma_2^2}}
\end{equation}

where $\mathbf{n}$ is the statistical number of $7_{th}$NND, $\mathbf{d}_{k,nn}$ is the distance, the fitted range of $\mathbf{d}_{k,nn}$ is (0, $\mathbf{d}_{k,nn}$ [n=max (n)]), and $(a, \mu, \sigma)_i$ are coefficients of the Gaussian functions. As shown in Fig.~\ref{fig1}, we used two functions (curves 1 and 2) obtained by Eq.~\ref{eqn1} to fit the $7_{th}$NND histogram. For a data set that contains any contributions from OCs, we found that the two Gaussian functions have a real number solution (e.g. the vertical red line in Fig.~\ref{fig1}), and it is set to be the $\epsilon$ value used in DBSCAN.

\end{itemize}


\subsection{Isochrone fitting}\label{cmd}
To estimate the physical parameters of the cluster candidates, we fit isochrones to the CMDs derived from the member stars of the clusters. The isochrones used here contain logarithmic ages from 5.92 to 10.12 with a step of 0.02, and metal fractions from 0.015 to 0.029 with a step of 0.001
\footnotemark[1]\footnotetext[1]{
  \url{http://stev.oapd.inaf.it/cgi-bin/cmd}},
 which come from the PARSEC library \citep{Bressan12} and have been updated for the \emph{Gaia} DR2 passbands using the photometric calibrations of \citet{Evans18}. The isochrones were corrected for extinction and reddening, using an extinction curve with $R_{\rm V}$ = 3.1 \citep{Cardelli89,Donnell94}. 

We selected clusters whose CMDs appeared to best-fit the theoretical isochrones. Following the method of \citet{Liu19}, the mean square distances to the isochrones were used to match the observed cluster members using Eq.~\ref{eq:opt_func}:

\begin{equation}\label{eq:opt_func}
\bar{d^2}= \frac{\sum_{k=1}^{n}(\mathbf{x}_k-\mathbf{x}_{k, nn})^2}{n}
\end{equation}

where $n$ is the number of core members (Section~\ref{dbscan}) in a cluster, and $\mathbf{x}$, $\mathbf{x}_{k, nn}$ are the positions of the member stars and the points on the isochrone that are closest to the member stars, respectively. At the same time, we also obtained the standard deviation of the square distance, $\sigma_{d^2}$, which was used to reflect the dispersion of the core members along the isochrones (Section~\ref{newcatlog}).

\section{Results and discussion}\label{results}

We applied the algorithm described in Section~\ref{dbscan} to about 170 million stars.  Since some clusters were located near the edges of each rectangle (Section~\ref{preparing}), in order not to find duplicate clusters or substructures of known clusters, we merged the cluster results within a range of 3$\sigma$ of the astrometric data $(l, b, \varpi, \mu_{\alpha^*}, \mu_{\delta})$; meanwhile, a visual inspection was also adopted. Moreover, to reduce the influence of outliers, we omitted clustering results with $n_{core}$ < 5. After that, we obtained 3,066 groups with a total of 371,362 members, of which 261,377 were core members. Thereafter, we removed all known OCs from this sample (Section \ref{cross}).

\subsection{Cross matching with previous catalogues}\label{cross}

\subsubsection{OCs in \emph{Gaia} DR2}
Our first cross-matched catalogue contained OCs derived from \emph{Gaia} DR2. In the previous studies, \citet{Sim19} found 207 new OCs within 1 kpc of solar system, while \citet[][]{Liu19} found 2,443 star clusters and candidates, 76 of which were most probable new star clusters.
\citet{CG20} re-visited 2017 OCs, including the catalogues of \citet[][]{CG18,CG19-0,CG20_0,Castro18,Castro19,Castro20}, and a portion of OCs published by \citet[][]{Sim19,Liu19}. In addition, we considered the 28, 16 and four new OCs recently been found by \citet{Ferreira19,Ferreira20,Hao20,Qin20}, respectively.

Most OCs are observed to have projected spatial extents up to nearly 15 pc~\citep{Gaia17} from the cluster centres, such as the apparent radii of OCs found in \emph{Gaia} DR2 by \citet{CG20_0}. Therefore, within the range of the maximum distance from the centre of an OC considered here (15 pc, 5$\sigma$ dispersion) within $(l, b)$, we extracted the reported clusters adjacent to the center of each identified group. Then, in the range of a 5$\sigma$ dispersion about $(\varpi, \mu_{\alpha^*},\mu_{\delta})$, the extracted clusters were matched, and the matched ones were removed. Finally, a total of 1,978 reported star clusters were removed.

\subsubsection{OCs before \emph{Gaia}}

We continue to match the remaining groups with the catalogues published by \citet[][D02]{Dias02} and \citet[][K13]{Kharchenko13}. To do this we made use of the root mean square (RMS) differences of the ages and distance moduli between the K13 clusters and cross-matched clusters identified in \emph{Gaia} DR2 by \citet{CG20}, which reflected the differences of these parameters between the re-identified clusters in \emph{Gaia} DR2 and the previous known OCs. We selected clusters in a range of max(15 pc, 5$\sigma$ dispersion) about $(l, b)$, and then compared the ages and distance moduli of the remaining groups with those of the K13 and D02 clusters within a 2-RMS level. In all, 46 groups were removed.

Next, \citet[][B19]{Bica19} constructed a sample 4,384 open/globular clusters, cluster candidates and cluster remnants, where most of them were previously catalogued by D02 and K13. We also matched the objects found by B19 with the remaining groups, where the radius of each matched cluster had to satisfy $d_r \leq radius \leq r_{15pc}$, where $d_r$ is the angular size between the matched group and the reported cluster. Although most star clusters could be physically distinguished, we also performed visual checks of the groups and reported clusters. In the end, we found that some groups were closely matched with the centre positions catalogued by B19, and we included two of them in the final table (Table~\ref{tab1}). After all the checking and filtering, 986 groups remained in our collated sample.

\subsection{New cluster candidates} \label{newcatlog}

\begin{figure*}
\begin{center}
	\includegraphics[width=1.\linewidth]{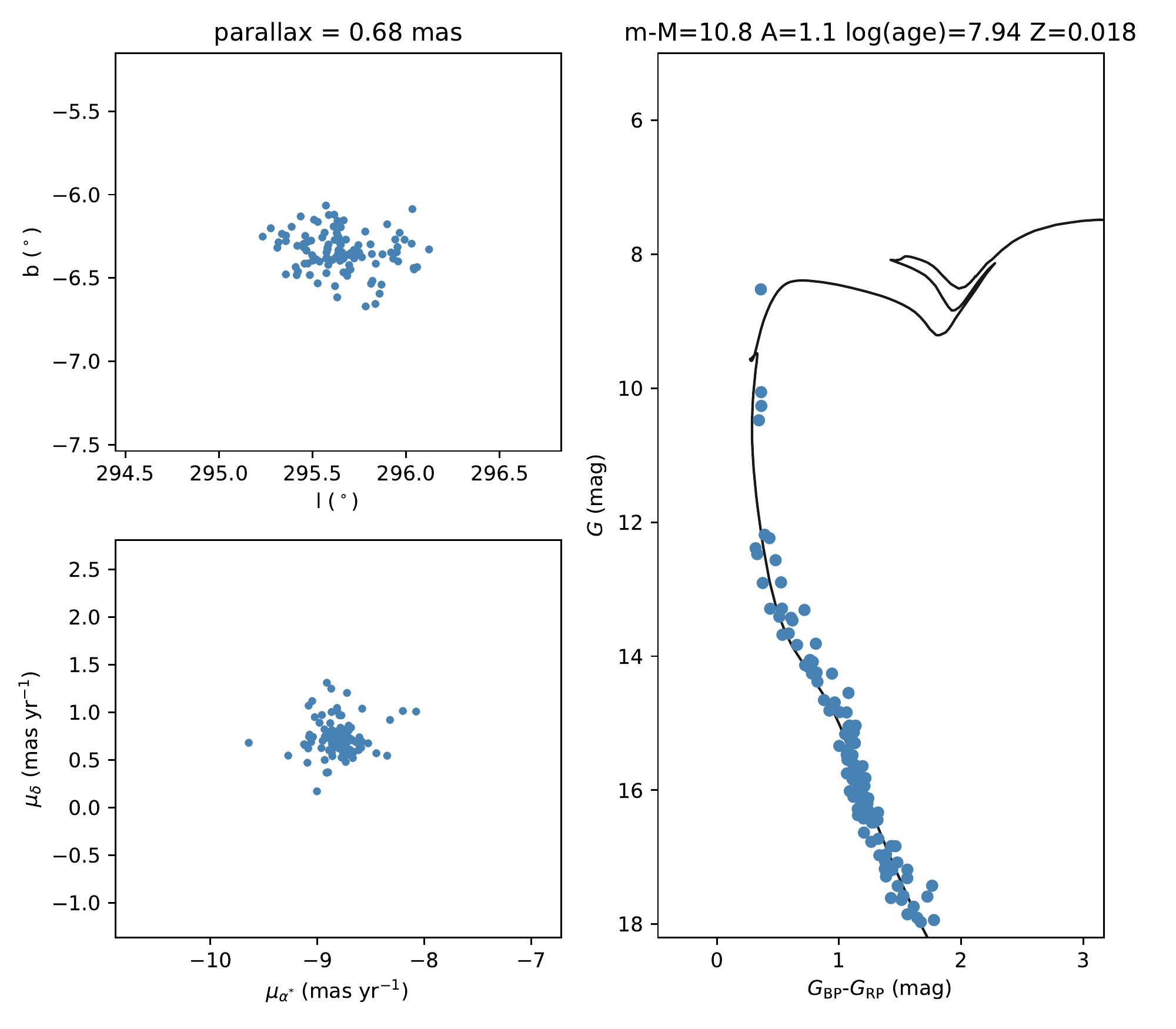}
	\caption{
	Example position and proper motion distributions, and CMDs of the member stars in a cluster candidate (NO.016 in Table~\ref{tab1}). The black line presents the best-fitting isochrone.
	}
	\label{fig2}
\end{center}
\end{figure*}

To find any probable new star clusters among the 986 groups, we selected clusters whose CMDs coincided well with the stellar isochrones (Section ~\ref{cmd}). Inspired by \citet{Liu19}, we first adopted criteria of
$n > 30; \bar{d^2} < 0.02; \sigma_{d^2} < 0.04$. After application of these criteria, 168 clusters with well-fitting isochrones remained in the sample. Furthermore, we performed visual inspections under careful consideration of the distributions of the positions and proper motions, and the CMDs of the remaining clusters. After that, 74 OC candidates remained. In Table~\ref{tab1}, we have presented the locations and sizes of these candidates, including their parallaxes, proper motions and corresponding 1$\sigma$ dispersions, which have also been presented in the works of \citet[][]{CG20_0,CG20} in the same way. The distance moduli, ages, extinction and metal fraction values of these candidates are also provided. The distributions, proper motions, CMDs (e.g. Fig.~\ref{fig2}) and member stars of each cluster are also shown in electric form.

\begin{center}
\setlength{\tabcolsep}{1mm}
\begin{longtable}{lccccccccccr}
	\caption{Parameters of the newly identified OC candidates.}
	\label{tab1}\\
	\toprule
NO.&$l$             &$b$     &$\theta$        &$n (n_{core}$)&$\varpi (\sigma_{\varpi})$&$\mu_{\alpha^{*}} (\sigma_{\mu_{\alpha^{*}}})$& $\mu_{\delta} (\sigma_{\mu_{\delta}})$&$m-M$ &log(age/yr)& $A_G$ &$Z$\\
	
          &($^{\circ}$)&($^{\circ}$) &(deg) &   &(mas)    &(mas~  yr$^{-1}$)     &(mas~  yr$^{-1}$)&  (mag)    &  & (mag) &\\
\midrule
001&94.203&-5.840& 0.20& 45(20)& 0.67(0.03)& 1.28(0.16)&-2.34(0.12)& 10.9& 8.40&  1.0&0.016\\
002&215.533&4.061& 0.11& 44(22)& 0.50(0.05)&-0.51(0.11)&-1.48(0.11)& 11.5& 8.12&  0.9&0.025\\
003&237.912$^{a}$&0.449& 0.18&143(75)& 0.55(0.05)&-1.38(0.13)& 1.24(0.13)& 11.4& 8.36&  0.6&0.026\\
004&116.192&-1.297& 0.09& 32(15)& 0.37(0.03)&-1.78(0.09)&-1.02(0.06)& 12.0& 7.82&  2.2&0.021\\
005&228.825&1.196& 0.26& 32(15)& 1.18(0.03)&-1.41(0.18)& 0.44(0.26)&  9.7& 8.04&  0.4&0.028\\
006&354.888&-1.335& 0.14& 70(45)& 1.05(0.05)& 1.12(0.12)&-0.09(0.11)&  9.3& 8.84&  0.3&0.025\\
007&11.026&-1.960& 0.13& 52(25)& 0.79(0.05)&-0.06(0.13)&-1.62(0.12)& 10.5& 7.48&  0.9&0.024\\
008&87.492$^{a}$&-3.675& 0.37& 50(24)& 0.85(0.04)& 0.05(0.14)&-4.35(0.13)& 10.4& 8.54&  1.0&0.027\\
009&241.773&0.143& 0.17& 41(13)& 0.38(0.03)&-1.63(0.09)& 1.31(0.12)& 11.9& 7.98&  0.6&0.026\\
010&339.546&-4.429& 0.22& 67(45)& 1.09(0.05)&-1.33(0.14)&-2.67(0.15)&  9.9& 7.98&  1.0&0.027\\
011&261.833&-9.192& 0.43&101(39)& 0.63(0.04)&-3.79(0.18)& 4.92(0.18)& 11.3& 7.44&  0.9&0.024\\
012&202.447&-5.870& 0.23& 54(28)& 0.86(0.06)&-0.01(0.18)&-1.43(0.17)& 10.4& 7.58&  0.9&0.021\\
013&315.674&1.631& 0.25& 57(36)& 0.87(0.04)&-1.20(0.10)&-3.11(0.13)& 10.3& 7.54&  1.1&0.017\\
014&314.563&2.660& 0.15& 46(16)& 0.52(0.02)&-2.66(0.12)&-2.72(0.10)& 11.3& 8.52&  0.8&0.027\\
015&259.105&-2.451& 0.24& 81(41)& 0.70(0.04)&-5.76(0.14)& 4.21(0.14)& 10.9& 8.02&  1.1&0.027\\
016&295.638&-6.345& 0.21&107(63)& 0.68(0.04)&-8.80(0.13)& 0.72(0.14)& 10.8& 7.94&  1.1&0.018\\
017&226.571&0.150& 0.24& 43(21)& 0.93(0.07)&-2.11(0.19)&-0.18(0.13)& 10.3& 8.42&  0.7&0.024\\
018&165.573&-2.123& 0.30& 46(19)& 0.95(0.07)&-0.40(0.17)&-3.05(0.14)&  9.9& 7.54&  0.9&0.019\\
019&210.343&-4.082& 0.14& 41(24)& 0.76(0.05)& 1.44(0.10)&-1.31(0.10)& 10.6& 8.60&  1.5&0.017\\
020&303.139&-0.852& 0.09& 59(28)& 0.63(0.07)&-6.34(0.19)&-1.58(0.14)& 10.6& 7.94&  1.7&0.024\\
021&224.698&3.430& 0.11& 46(19)& 0.46(0.06)&-0.80(0.12)&-0.87(0.11)& 11.1& 9.02&  0.1&0.019\\
022&224.178&1.248& 0.22& 47(30)& 0.86(0.06)&-0.57(0.11)& 0.83(0.20)& 10.5& 7.76&  0.7&0.016\\
023&102.341&-2.235& 0.10& 38(26)& 0.68(0.04)& 1.71(0.09)&-2.25(0.13)& 11.0& 8.72&  1.1&0.028\\
024&236.831&2.172& 0.10& 39(27)& 0.59(0.05)&-0.99(0.07)& 0.59(0.10)& 11.2& 7.96&  1.0&0.025\\
025&312.078&-2.372& 0.06& 34(14)& 0.48(0.03)&-4.72(0.08)&-3.13(0.15)& 11.2& 7.60&  1.2&0.021\\
026&240.457&-4.731& 0.31& 85(55)& 0.80(0.05)&-4.29(0.13)& 3.85(0.11)& 10.5& 7.44&  0.1&0.026\\
027&333.864&-4.275& 0.13& 32(13)& 0.59(0.01)&-1.58(0.14)&-4.01(0.07)& 11.1& 8.66&  1.1&0.028\\
028&100.808&0.620& 0.14& 88(42)& 0.64(0.05)&-2.01(0.16)&-1.63(0.12)& 10.8& 7.92&  2.2&0.027\\
029&2.373&-0.261& 0.22& 63(45)& 1.66(0.05)& 0.89(0.19)&-5.12(0.18)&  9.0& 7.70&  0.8&0.023\\
030&24.838&-6.092& 0.22& 43(27)& 0.94(0.04)& 1.13(0.11)& 0.84(0.10)& 10.3& 8.50&  1.1&0.026\\
031&86.396&-0.476& 0.04& 33(16)& 0.49(0.03)&-2.51(0.13)&-4.52(0.14)& 11.0& 8.14&  1.7&0.020\\
032&221.101&-1.049& 0.19& 51(30)& 0.88(0.04)&-1.59(0.13)&-3.46(0.14)& 10.5& 8.54&  0.8&0.025\\
033&343.054&2.668& 0.18&100(56)& 0.90(0.05)& 1.42(0.19)&-2.98(0.16)& 10.1& 7.64&  1.2&0.016\\
034&309.876&1.271& 0.14& 41(27)& 0.75(0.04)&-7.20(0.09)&-2.26(0.11)& 10.4& 7.98&  0.8&0.025\\
035&161.079&-0.654& 0.09& 34(17)& 0.52(0.05)& 1.43(0.13)&-2.11(0.26)& 11.5& 8.44&  1.0&0.024\\
036&263.281&-6.758& 0.27& 32(15)& 0.93(0.04)&-5.61(0.15)& 5.11(0.15)& 10.1& 7.88&  0.4&0.027\\
037&20.457&-0.866& 0.15& 37(19)& 0.92(0.05)&-0.06(0.14)&-3.05(0.12)&  9.9& 8.82&  1.4&0.023\\
038&178.173&2.255& 0.15& 30(14)& 0.71(0.05)& 0.49(0.15)&-3.81(0.10)& 10.8& 8.46&  0.7&0.026\\
039&121.599&6.001& 0.33& 37(21)& 0.96(0.03)&-1.53(0.14)&-0.57(0.11)&  9.5& 7.94&  1.9&0.016\\
040&70.424&-1.231& 0.03& 50(38)& 0.48(0.03)&-3.83(0.09)&-5.30(0.14)& 11.1& 8.24&  3.9&0.024\\
041&92.224&-6.323& 0.61& 68(29)& 0.78(0.04)& 2.15(0.24)&-0.02(0.11)& 10.2& 8.88&  0.4&0.017\\
042&252.836&-1.370& 0.31& 66(36)& 0.86(0.04)&-4.78(0.10)& 2.97(0.15)& 10.4& 7.96&  0.7&0.026\\
043&237.960&0.637& 0.19& 90(41)& 0.66(0.04)&-2.52(0.18)& 1.65(0.12)& 10.9& 8.46&  0.3&0.026\\
044&134.977&1.725& 0.18& 34(20)& 0.67(0.03)&-0.91(0.18)&-0.46(0.13)& 10.8& 7.66&  1.8&0.017\\
045&111.351&-2.453& 0.14& 59(40)& 0.99(0.04)&-0.42(0.12)&-0.83(0.12)&  9.9& 7.90&  2.2&0.020\\
046&68.596&2.636& 0.09& 56(37)& 0.49(0.04)&-2.51(0.10)&-3.43(0.11)& 11.3& 8.34&  1.7&0.018\\
047&327.232&-1.331& 0.07& 33(11)& 0.57(0.03)&-3.95(0.16)&-5.16(0.11)& 10.6& 8.78&  2.0&0.024\\
048&225.173&5.548& 0.15& 33(17)& 0.90(0.08)&-3.59(0.14)& 0.99(0.11)& 10.2& 7.98&  0.1&0.026\\
049&169.769&-1.966& 0.13& 33(17)& 0.60(0.04)& 1.15(0.13)&-3.97(0.09)& 10.9& 8.18&  1.2&0.024\\
050&203.158&-2.565& 0.07& 41(20)& 0.49(0.05)&-0.90(0.16)&-0.24(0.10)& 11.7& 8.16&  1.0&0.019\\
051&19.445&-2.247& 0.08& 45(23)& 0.56(0.04)& 0.67(0.11)&-0.14(0.11)& 10.9& 8.10&  1.9&0.025\\
052&220.279&0.592& 0.17& 35(20)& 0.60(0.04)&-0.26(0.11)&-0.64(0.16)& 10.7& 7.92&  0.5&0.021\\
053&174.786&-1.207& 0.14& 35(13)& 0.64(0.06)& 1.41(0.17)&-3.54(0.18)& 11.2& 7.54&  1.1&0.023\\
054&221.571&0.052& 0.16& 50(33)& 0.93(0.02)&-3.94(0.15)& 1.14(0.14)& 10.3& 8.42&  0.4&0.028\\
055&298.570&-8.573& 0.35& 42(18)& 0.73(0.05)&-4.05(0.16)&-0.69(0.13)& 10.5& 7.78&  0.9&0.017\\
056&92.545&-3.423& 0.10& 34(18)& 0.78(0.05)& 1.44(0.08)&-0.35(0.11)& 10.6& 7.90&  0.9&0.027\\
057&16.639&-1.919& 0.13& 40(19)& 0.86(0.05)&-2.70(0.08)&-3.56(0.08)& 10.1& 8.48&  0.7&0.028\\
058&203.355&0.556& 0.12& 38(15)& 0.48(0.04)& 0.03(0.11)&-1.38(0.07)& 11.3& 7.28&  1.1&0.018\\
059&235.209&1.629& 0.06& 35(20)& 0.41(0.04)&-2.23(0.10)& 2.19(0.09)& 12.0& 7.52&  1.5&0.021\\
060&210.929&5.473& 0.13& 55(29)& 0.48(0.04)&-0.39(0.11)&-1.66(0.12)& 11.5& 8.30&  0.9&0.022\\
061&221.759&-4.652& 0.14& 73(37)& 0.50(0.04)&-1.28(0.12)& 1.25(0.15)& 11.6& 8.28&  1.4&0.020\\
062&233.570&-4.761& 0.10& 36(14)& 0.49(0.05)&-1.60(0.14)& 0.67(0.13)& 11.3& 8.38&  1.6&0.024\\
063&133.389&-0.837& 0.08& 39(21)& 0.63(0.06)&-0.89(0.09)& 0.16(0.18)& 10.8& 8.00&  2.1&0.022\\
064&293.050&-6.779& 0.28& 46(27)& 0.93(0.04)&-7.41(0.14)& 1.72(0.19)& 10.2& 7.90&  1.0&0.022\\
065&245.796&3.234& 0.19& 32(23)& 0.99(0.04)&-3.93(0.11)& 2.00(0.10)& 10.1& 8.20&  0.2&0.028\\
066&14.837&1.394& 0.05& 33(13)& 0.61(0.03)&-0.25(0.12)&-1.73(0.15)& 11.4& 6.80&  1.4&0.028\\
067&323.479&-2.512& 0.06& 33(16)& 0.54(0.03)&-2.17(0.07)&-3.52(0.11)& 10.7& 8.38&  1.8&0.028\\
068&262.909&-4.363& 0.38& 71(31)& 0.82(0.05)&-6.29(0.20)& 5.02(0.33)& 10.3& 7.78&  0.7&0.017\\
069&234.158&-12.181& 0.33& 93(66)& 0.94(0.06)&-1.52(0.12)& 2.45(0.15)& 10.1& 8.60&  0.4&0.018\\
070&254.535&-6.050& 0.28& 31(11)& 0.82(0.05)&-4.06(0.16)& 4.59(0.21)& 10.3& 7.98&  0.6&0.027\\
071&255.762&-4.530& 0.14& 36(18)& 0.67(0.04)&-4.20(0.14)& 3.38(0.08)& 11.1& 8.52&  1.2&0.028\\
072&217.519&0.474& 0.22& 35(12)& 0.64(0.04)&-0.15(0.09)&-0.80(0.10)& 11.3& 7.86&  0.5&0.028\\
073&147.486&4.112& 0.15& 68(36)& 0.69(0.06)& 1.04(0.10)&-1.06(0.15)& 10.4& 7.82&  2.1&0.018\\
074&216.760&-0.834& 0.05& 35(20)& 0.42(0.05)&-1.81(0.11)& 1.07(0.07)& 11.6& 8.76&  1.2&0.019\\
\bottomrule
\end{longtable}
 \begin{tablenotes}
        \footnotesize
       \item$^{a}$ Adjacent to the cluster catalogued by \citet{Bica19}.
      \end{tablenotes}
      
\end{center}

\begin{figure*}
\begin{center}
	\includegraphics[width=1.\linewidth]{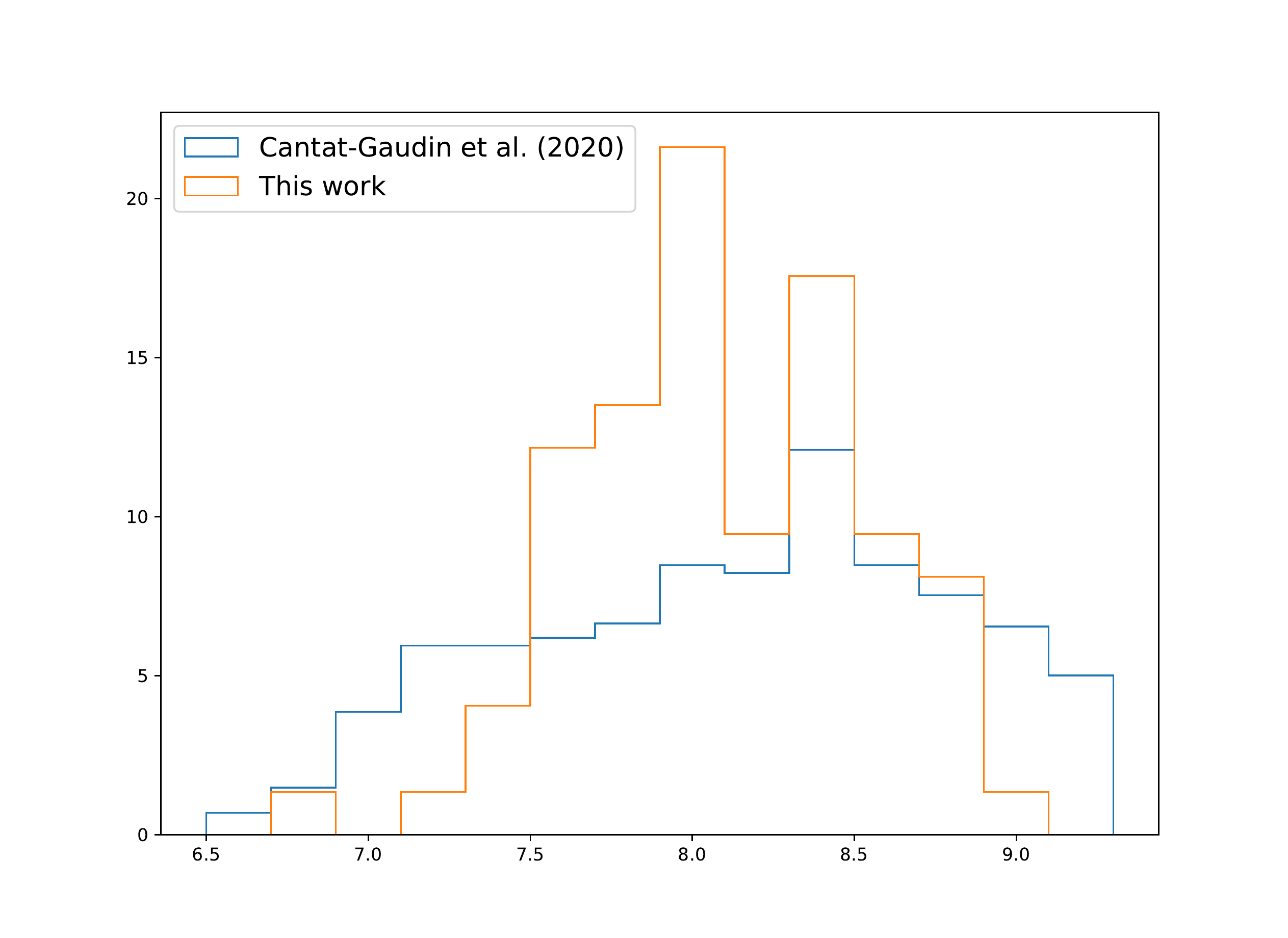}
	\caption{
	Frequency distribution of the logarithmic ages of the cluster candidates derived in this study, which have been compared to those of \emph{Gaia} DR2 confirmed OCs \citep{CG20}.
	}
	\label{fig3}
\end{center}
\end{figure*}

\begin{figure*}
\begin{center}
	\includegraphics[width=1.\linewidth]{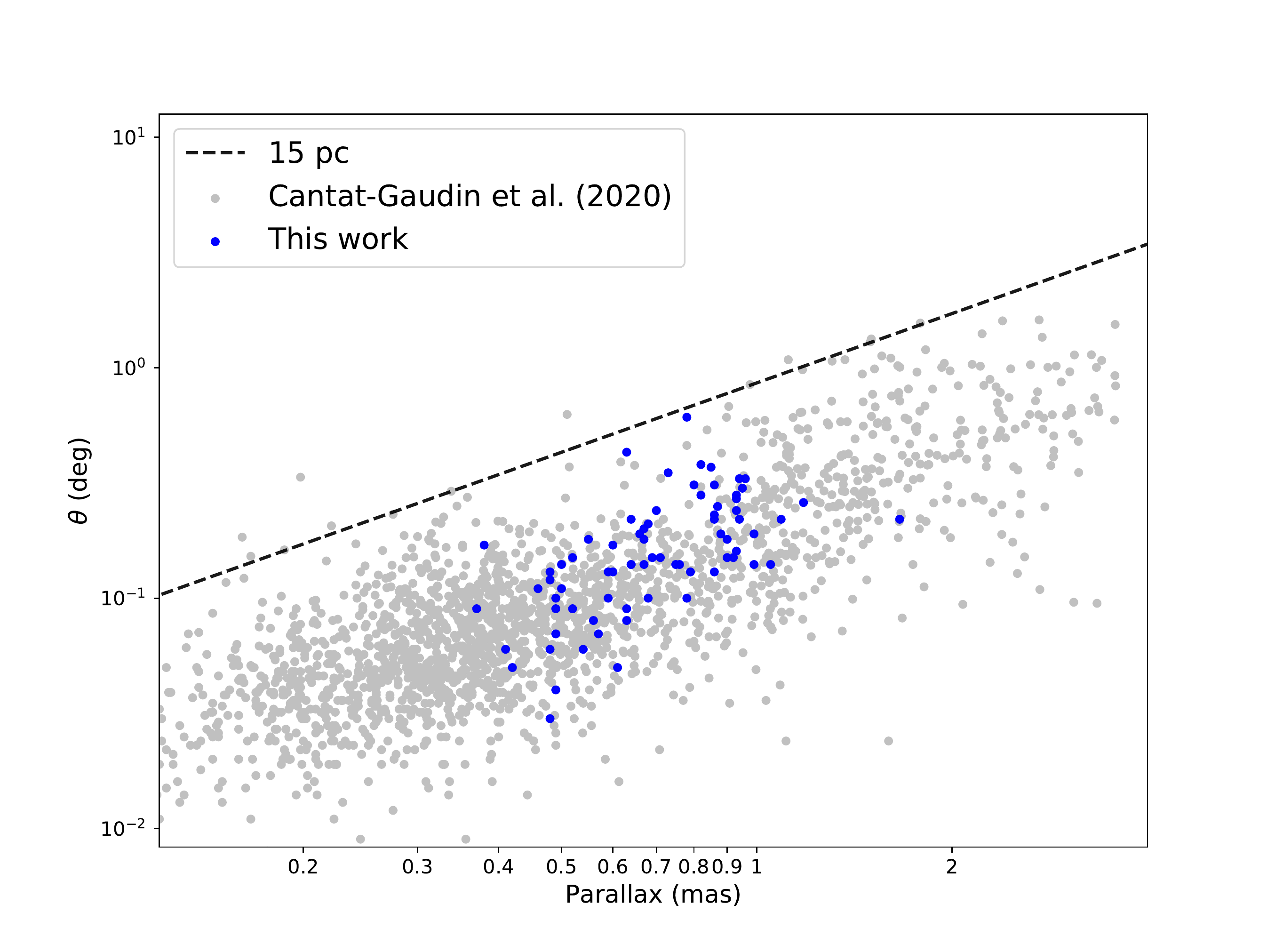}
	\caption{
	Apparent radius as a function of parallax for the cluster candidates discovered in this work (blue dots). The grey dots show the OCs cataloged by \citet{CG20} in \emph{Gaia} DR2. 	The black dashed line indicates an angular size corresponding to 15 pc \citep{CG20_0}.
	}
	\label{fig4}
\end{center}
\end{figure*}

As shown in Fig.~\ref{fig3}, the isochrones show that the ages of most of the new OC candidates are in the range of log (age/yr) = (7.5, 8.5), and two extreme values are located at log (age/yr) = 8.0 and 8.4. These results are similar to the age distributions of the OCs presented by \citet[][CG20]{CG20}. However, the peak value of log (age/yr) = 8.0 may be due to bias caused by the new cluster candidates found in this work forming an incomplete sample. In addition, as shown in Fig.~\ref{fig4}, the apparent radii of the new star cluster candidates are below 15 pc, which is consistent with the distributions of apparent radii of OCs catalogued by CG20.

\begin{figure*}
\begin{center}
	\includegraphics[width=1.\linewidth]{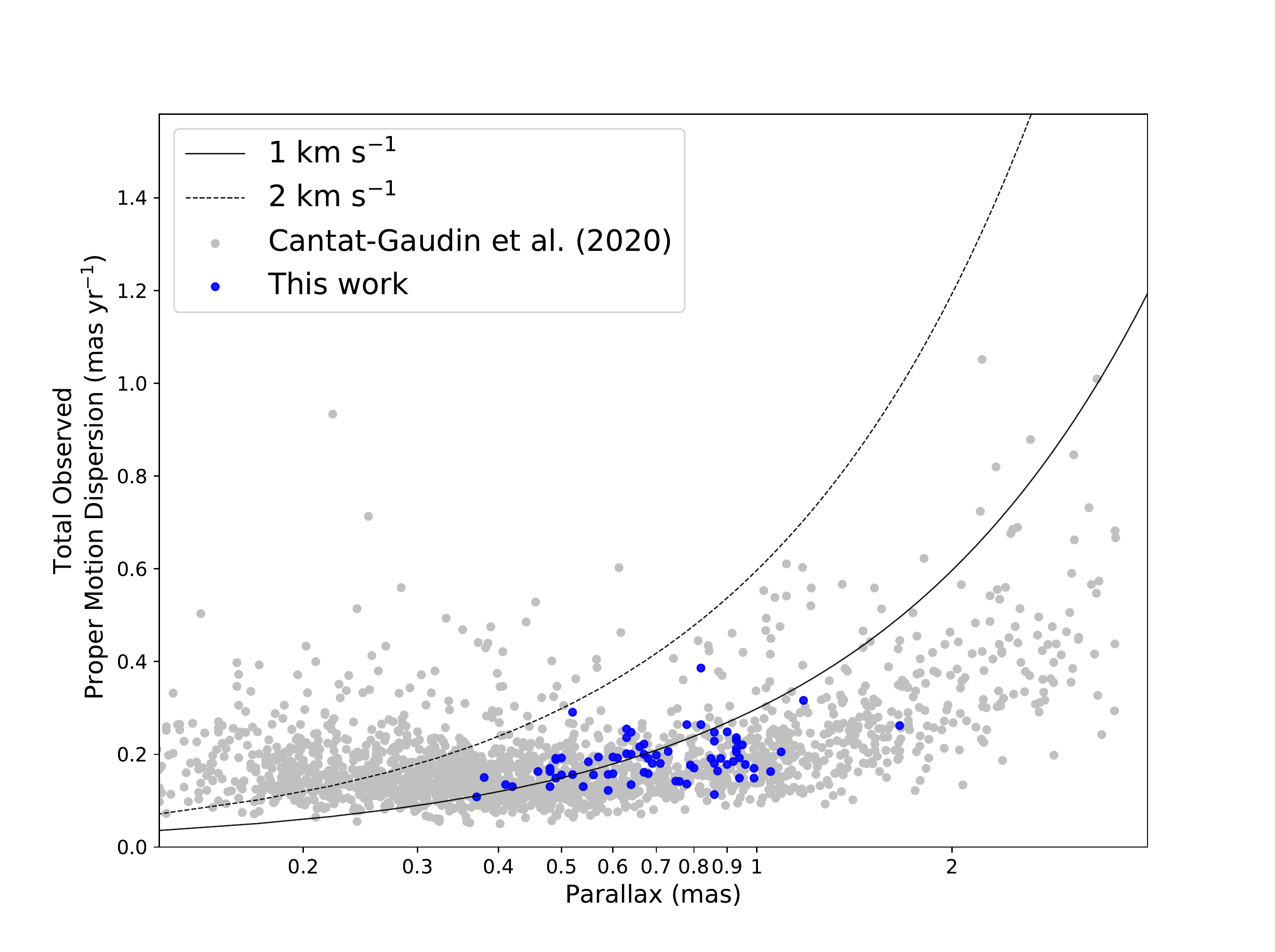}
	\caption{Total proper motion dispersion as a function of parallax for the cluster candidates discovered in this work (blue dots). The grey dots show the OCs cataloged by \citet{CG20} in \emph{Gaia} DR2. The black solid and dashed lines show the proper motion dispersions corresponding to theoretical 1-D velocity dispersions of 1 and 2~\kms, respectively, in the absence of any measurement errors \citep{CG20_0}.
	}
	\label{fig5}
\end{center}
\end{figure*}

As a gravitationally bound system, member stars of an OC have much smaller internal velocity dispersions than field stars. Radial velocities derived from high-resolution spectroscopic observations have shown that the internal velocity dispersion of OCs are generally less than 2~\kms   \citep[e.g.][]{Donati14,CG14,Vereshchagin16,Overbeek17}. Meanwhile, the observed proper motion dispersions of the clusters found in \emph{Gaia} DR2 are correlated with their parallaxes \citep{CG20_0,Ferreira20}. As shown in Fig.~\ref{fig5}, the proper motion dispersions of the new star cluster candidates found in this work are below 0.5~mas~yr$^{-1}$, and the resolved velocity dispersions of most candidates are below 1~\kms, which are consistent with the results of star clusters found in \emph{Gaia} DR2 by \citet{CG20}.

\section{Conclusion}\label{sec:summary}
Adopting the clustering algorithm DBSCAN, we used the positions, parallaxes, and proper motions of nearly 170 million stars within the galactic latitude range $|b| < 20^{\circ}$ and parallax $\varpi >$ 0.2~mas in \emph{Gaia} DR2 to search for unknown OCs. We cross-matched our initial sample with previous star cluster catalogues, and after performing several vetting stages, we obtained a catalogue of 74 new OC candidates. We fitted isochrones to these candidates and obtained their corresponding physical parameters. The apparent radii and observed proper motion dispersions of the new cluster candidates were consistent with those of OCs previously found in \emph{Gaia} DR2. 

The detection of new OC candidates shows that there are still some missed OCs in the \emph{Gaia} data. \emph{Gaia} EDR3 will be initially released in December 2020, with the complete release due the first half of 2022, which will contain higher precision parallaxes, proper motions and photometric data. Therefore, the newly discovered OC candidates in this work can be further confirmed their natures, and more star clusters are expected to be found in future works.

\section{Acknowledgements}
This work has made use of data from the European Space Agency (ESA)
mission
 \emph{Gaia}(\url{https://www.cosmos.esa.int/Gaia}), processed by the  \emph{Gaia} Data Processing and Analysis Consortium (DPAC,
\url{https://www.cosmos.esa.int/web/Gaia/dpac/consortium}). Funding for the DPAC has been provided by national institutions, in particular the institutions participating in the \emph{Gaia} Multilateral Agreement.
This work was funded by the NSFC (grant numbers 11933011, 11873019, and 11673066), and by the Key Laboratory for Radio Astronomy.
This research has made use of the open-source Python packages \texttt{Astropy} \citep{astropy}, \texttt{NumPy} \citep{numpy}, \texttt{scikit-learn} \citep{Pedregosa11} and \texttt{Pandas} \citep{pandas}. The figures in this article were created using \texttt{Matplotlib} \citep{hunter07}. 

\bibliographystyle{aasjournal} 
\bibliography{noc} 

\begin{thebibliography}{}
\expandafter\ifx\csname natexlab\endcsname\relax\def\natexlab#1{#1}\fi

\bibitem[{{Bica} {et~al.}(2019){Bica}, {Pavani}, {Bonatto}, \& {Lima}}]{Bica19}
{Bica}, E., {Pavani}, D.~B., {Bonatto}, C.~J., \& {Lima}, E.~F. 2019, \aj, 157,
  12

\bibitem[{{Bressan} {et~al.}(2012){Bressan}, {Marigo}, {Girardi}, {Salasnich},
  {Dal Cero}, {Rubele}, \& {Nanni}}]{Bressan12}
{Bressan}, A., {Marigo}, P., {Girardi}, L., {et~al.} 2012, \mnras, 427, 127

\bibitem[{{Cantat-Gaudin} \& {Anders}(2020)}]{CG20_0}
{Cantat-Gaudin}, T., \& {Anders}, F. 2020, \aap, 633, A99

\bibitem[{{Cantat-Gaudin} {et~al.}(2014){Cantat-Gaudin}, {Vallenari}, {Zaggia},
  {Bragaglia}, {Sordo}, {Drew}, {Eisloeffel}, {Farnhill}, {Gonzalez-Solares},
  {Greimel}, {Irwin}, {Kupcu-Yoldas}, {Jordi}, {Blomme}, {Sampedro}, {Costado},
  {Alfaro}, {Smiljanic}, {Magrini}, {Donati}, {Friel}, {Jacobson}, {Abbas},
  {Hatzidimitriou}, {Spagna}, {Vecchiato}, {Balaguer-Nunez}, {Lardo}, {Tosi},
  {Pancino}, {Klutsch}, {Tautvaisiene}, {Drazdauskas}, {Puzeras},
  {Jim{\'e}nez-Esteban}, {Maiorca}, {Geisler}, {San Roman}, {Villanova},
  {Gilmore}, {Rand ich}, {Bensby}, {Flaccomio}, {Lanzafame}, {Recio-Blanco},
  {Damiani}, {Hourihane}, {Jofr{\'e}}, {de Laverny}, {Masseron}, {Morbidelli},
  {Prisinzano}, {Sacco}, {Sbordone}, \& {Worley}}]{CG14}
{Cantat-Gaudin}, T., {Vallenari}, A., {Zaggia}, S., {et~al.} 2014, \aap, 569,
  A17

\bibitem[{{Cantat-Gaudin} {et~al.}(2018){Cantat-Gaudin}, {Jordi}, {Vallenari},
  {Bragaglia}, {Balaguer-N{\'u}{\~n}ez}, {Soubiran}, {Bossini}, {Moitinho},
  {Castro-Ginard}, {Krone-Martins}, {Casamiquela}, {Sordo}, \&
  {Carrera}}]{CG18}
{Cantat-Gaudin}, T., {Jordi}, C., {Vallenari}, A., {et~al.} 2018, \aap, 618,
  A93

\bibitem[{{Cantat-Gaudin} {et~al.}(2019){Cantat-Gaudin}, {Krone-Martins},
  {Sedaghat}, {Farahi}, {de Souza}, {Skalidis}, {Malz}, {Mac{\^e}do}, {Moews},
  {Jordi}, {Moitinho}, {Castro-Ginard}, {Ishida}, {Heneka}, {Boucaud}, \&
  {Trindade}}]{CG19-0}
{Cantat-Gaudin}, T., {Krone-Martins}, A., {Sedaghat}, N., {et~al.} 2019, \aap,
  624, A126

\bibitem[{{Cantat-Gaudin} {et~al.}(2020){Cantat-Gaudin}, {Anders},
  {Castro-Ginard}, {Jordi}, {Romero-G{\'o}mez}, {Soubiran}, {Casamiquela},
  {Tarricq}, {Moitinho}, {Vallenari}, {Bragaglia}, {Krone-Martins}, \&
  {Kounkel}}]{CG20}
{Cantat-Gaudin}, T., {Anders}, F., {Castro-Ginard}, A., {et~al.} 2020, \aap,
  640, A1

\bibitem[{{Cardelli} {et~al.}(1989){Cardelli}, {Clayton}, \&
  {Mathis}}]{Cardelli89}
{Cardelli}, J.~A., {Clayton}, G.~C., \& {Mathis}, J.~S. 1989, \apj, 345, 245

\bibitem[{{Carraro} {et~al.}(2007){Carraro}, {Geisler}, {Villanova},
  {Frinchaboy}, \& {Majewski}}]{Carraro07}
{Carraro}, G., {Geisler}, D., {Villanova}, S., {Frinchaboy}, P.~M., \&
  {Majewski}, S.~R. 2007, \aap, 476, 217

\bibitem[{{Castro-Ginard} {et~al.}(2019){Castro-Ginard}, {Jordi}, {Luri},
  {Cantat-Gaudin}, \& {Balaguer-N{\'u}{\~n}ez}}]{Castro19}
{Castro-Ginard}, A., {Jordi}, C., {Luri}, X., {Cantat-Gaudin}, T., \&
  {Balaguer-N{\'u}{\~n}ez}, L. 2019, \aap, 627, A35

\bibitem[{{Castro-Ginard} {et~al.}(2018){Castro-Ginard}, {Jordi}, {Luri},
  {Julbe}, {Morvan}, {Balaguer-N{\'u}{\~n}ez}, \& {Cantat-Gaudin}}]{Castro18}
{Castro-Ginard}, A., {Jordi}, C., {Luri}, X., {et~al.} 2018, \aap, 618, A59

\bibitem[{{Castro-Ginard} {et~al.}(2020){Castro-Ginard}, {Jordi}, {Luri},
  {{\'A}lvarez Cid-Fuentes}, {Casamiquela}, {Anders}, {Cantat-Gaudin},
  {Mongui{\'o}}, {Balaguer-N{\'u}{\~n}ez}, {Sol{\`a}}, \& {Badia}}]{Castro20}
---. 2020, \aap, 635, A45

\bibitem[{{Dias} {et~al.}(2002){Dias}, {Alessi}, {Moitinho}, \&
  {L{\'e}pine}}]{Dias02}
{Dias}, W.~S., {Alessi}, B.~S., {Moitinho}, A., \& {L{\'e}pine}, J.~R.~D. 2002,
  \aap, 389, 871

\bibitem[{{Donati} {et~al.}(2014){Donati}, {Cantat Gaudin}, {Bragaglia},
  {Friel}, {Magrini}, {Smiljanic}, {Vallenari}, {Tosi}, {Sordo},
  {Tautvai{\v{s}}ien{\.{e}}}, {Blanco-Cuaresma}, {Costado}, {Geisler},
  {Klutsch}, {Mowlavi}, {Mu{\~n}oz}, {San Roman}, {Zaggia}, {Gilmore},
  {Randich}, {Bensby}, {Flaccomio}, {Koposov}, {Korn}, {Pancino},
  {Recio-Blanco}, {Franciosini}, {de Laverny}, {Lewis}, {Morbidelli},
  {Prisinzano}, {Sacco}, {Worley}, {Hourihane}, {Jofr{\'e}}, {Lardo}, \&
  {Maiorca}}]{Donati14}
{Donati}, P., {Cantat Gaudin}, T., {Bragaglia}, A., {et~al.} 2014, \aap, 561,
  A94

\bibitem[{Ester {et~al.}(1996)Ester, Kriegel, Sander, \& Xu}]{Ester96}
Ester, M., Kriegel, H.-P., Sander, J., \& Xu, X. 1996, in Proc. of 2nd
  International Conference on Knowledge Discovery and Data Mining (KDD-96),
  226--231

\bibitem[{{Evans} {et~al.}(2018){Evans}, {Riello}, {De Angeli}, {Carrasco},
  {Montegriffo}, {Fabricius}, {Jordi}, {Palaversa}, {Diener}, {Busso},
  {Cacciari}, {van Leeuwen}, {Burgess}, {Davidson}, {Harrison}, {Hodgkin},
  {Pancino}, {Richards}, {Altavilla}, {Balaguer-N{\'u}{\~n}ez}, {Barstow},
  {Bellazzini}, {Brown}, {Castellani}, {Cocozza}, {De Luise}, {Delgado},
  {Ducourant}, {Galleti}, {Gilmore}, {Giuffrida}, {Holl}, {Kewley}, {Koposov},
  {Marinoni}, {Marrese}, {Osborne}, {Piersimoni}, {Portell}, {Pulone},
  {Ragaini}, {Sanna}, {Terrett}, {Walton}, {Wevers}, \&
  {Wyrzykowski}}]{Evans18}
{Evans}, D.~W., {Riello}, M., {De Angeli}, F., {et~al.} 2018, \aap, 616, A4

\bibitem[{{Ferreira} {et~al.}(2020){Ferreira}, {Corradi}, {Maia}, {Angelo}, \&
  {Santos}}]{Ferreira20}
{Ferreira}, F.~A., {Corradi}, W.~J.~B., {Maia}, F.~F.~S., {Angelo}, M.~S., \&
  {Santos}, J.~F.~C., J. 2020, \mnras, 496, 2021

\bibitem[{{Ferreira} {et~al.}(2019){Ferreira}, {Santos}, {Corradi}, {Maia}, \&
  {Angelo}}]{Ferreira19}
{Ferreira}, F.~A., {Santos}, J.~F.~C., {Corradi}, W.~J.~B., {Maia}, F.~F.~S.,
  \& {Angelo}, M.~S. 2019, \mnras, 483, 5508

\bibitem[{{Friel}(1995)}]{Friel95}
{Friel}, E.~D. 1995, \araa, 33, 381

\bibitem[{{Gaia Collaboration} {et~al.}(2016){Gaia Collaboration}, {Prusti},
  {de Bruijne}, {Brown}, {Vallenari}, {Babusiaux}, {Bailer-Jones}, {Bastian},
  {Biermann}, {Evans}, {Eyer}, {Jansen}, {Jordi}, {Klioner}, {Lammers},
  {Lindegren}, {Luri}, {Mignard}, {Milligan}, {Panem}, {Poinsignon},
  {Pourbaix}, {Randich}, {Sarri}, {Sartoretti}, {Siddiqui}, {Soubiran},
  {Valette}, {van Leeuwen}, {Walton}, {Aerts}, {Arenou}, {Cropper}, {Drimmel},
  {H{\o}g}, {Katz}, {Lattanzi}, {O'Mullane}, {Grebel}, {Holland}, {Huc},
  {Passot}, {Bramante}, {Cacciari}, {Casta{\~n}eda}, {Chaoul}, {Cheek}, {De
  Angeli}, {Fabricius}, {Guerra}, {Hern{\'a}ndez}, {Jean-Antoine-Piccolo},
  {Masana}, {Messineo}, {Mowlavi}, {Nienartowicz}, {Ord{\'o}{\~n}ez-Blanco},
  {Panuzzo}, {Portell}, {Richards}, {Riello}, {Seabroke}, {Tanga},
  {Th{\'e}venin}, {Torra}, {Els}, {Gracia-Abril}, {Comoretto},
  {Garcia-Reinaldos}, {Lock}, {Mercier}, {Altmann}, {Andrae}, {Astraatmadja},
  {Bellas-Velidis}, {Benson}, {Berthier}, {Blomme}, {Busso}, {Carry},
  {Cellino}, {Clementini}, {Cowell}, {Creevey}, {Cuypers}, {Davidson}, {De
  Ridder}, {de Torres}, {Delchambre}, {Dell'Oro}, {Ducourant}, {Fr{\'e}mat},
  {Garc{\'\i}a-Torres}, {Gosset}, {Halbwachs}, {Hambly}, {Harrison}, {Hauser},
  {Hestroffer}, {Hodgkin}, {Huckle}, {Hutton}, {Jasniewicz}, {Jordan},
  {Kontizas}, {Korn}, {Lanzafame}, {Manteiga}, {Moitinho}, {Muinonen},
  {Osinde}, {Pancino}, {Pauwels}, {Petit}, {Recio-Blanco}, {Robin}, {Sarro},
  {Siopis}, {Smith}, {Smith}, {Sozzetti}, {Thuillot}, {van Reeven}, {Viala},
  {Abbas}, {Abreu Aramburu}, {Accart}, {Aguado}, {Allan}, {Allasia},
  {Altavilla}, {{\'A}lvarez}, {Alves}, {Anderson}, {Andrei}, {Anglada Varela},
  {Antiche}, {Antoja}, {Ant{\'o}n}, {Arcay}, {Atzei}, {Ayache}, {Bach},
  {Baker}, {Balaguer-N{\'u}{\~n}ez}, {Barache}, {Barata}, {Barbier}, {Barblan},
  {Baroni}, {Barrado y Navascu{\'e}s}, {Barros}, {Barstow}, {Becciani},
  {Bellazzini}, {Bellei}, {Bello Garc{\'\i}a}, {Belokurov}, {Bendjoya},
  {Berihuete}, {Bianchi}, {Bienaym{\'e}}, {Billebaud}, {Blagorodnova},
  {Blanco-Cuaresma}, {Boch}, {Bombrun}, {Borrachero}, {Bouquillon}, {Bourda},
  {Bouy}, {Bragaglia}, {Breddels}, {Brouillet}, {Br{\"u}semeister},
  {Bucciarelli}, {Budnik}, {Burgess}, {Burgon}, {Burlacu}, {Busonero}, {Buzzi},
  {Caffau}, {Cambras}, {Campbell}, {Cancelliere}, {Cantat-Gaudin}, {Carlucci},
  {Carrasco}, {Castellani}, {Charlot}, {Charnas}, {Charvet}, {Chassat},
  {Chiavassa}, {Clotet}, {Cocozza}, {Collins}, {Collins}, {Costigan}, {Crifo},
  {Cross}, {Crosta}, {Crowley}, {Dafonte}, {Damerdji}, {Dapergolas}, {David},
  {David}, {De Cat}, {de Felice}, {de Laverny}, {De Luise}, {De March}, {de
  Martino}, {de Souza}, {Debosscher}, {del Pozo}, {Delbo}, {Delgado},
  {Delgado}, {di Marco}, {Di Matteo}, {Diakite}, {Distefano}, {Dolding}, {Dos
  Anjos}, {Drazinos}, {Dur{\'a}n}, {Dzigan}, {Ecale}, {Edvardsson}, {Enke},
  {Erdmann}, {Escolar}, {Espina}, {Evans}, {Eynard Bontemps}, {Fabre},
  {Fabrizio}, {Faigler}, {Falc{\~a}o}, {Farr{\`a}s Casas}, {Faye}, {Federici},
  {Fedorets}, {Fern{\'a}ndez-Hern{\'a}ndez}, {Fernique}, {Fienga}, {Figueras},
  {Filippi}, {Findeisen}, {Fonti}, {Fouesneau}, {Fraile}, {Fraser}, {Fuchs},
  {Furnell}, {Gai}, {Galleti}, {Galluccio}, {Garabato}, {Garc{\'\i}a-Sedano},
  {Gar{\'e}}, {Garofalo}, {Garralda}, {Gavras}, {Gerssen}, {Geyer}, {Gilmore},
  {Girona}, {Giuffrida}, {Gomes}, {Gonz{\'a}lez-Marcos},
  {Gonz{\'a}lez-N{\'u}{\~n}ez}, {Gonz{\'a}lez-Vidal}, {Granvik}, {Guerrier},
  {Guillout}, {Guiraud}, {G{\'u}rpide}, {Guti{\'e}rrez-S{\'a}nchez}, {Guy},
  {Haigron}, {Hatzidimitriou}, {Haywood}, {Heiter}, {Helmi}, {Hobbs},
  {Hofmann}, {Holl}, {Holland }, {Hunt}, {Hypki}, {Icardi}, {Irwin}, {Jevardat
  de Fombelle}, {Jofr{\'e}}, {Jonker}, {Jorissen}, {Julbe}, {Karampelas},
  {Kochoska}, {Kohley}, {Kolenberg}, {Kontizas}, {Koposov}, {Kordopatis},
  {Koubsky}, {Kowalczyk}, {Krone-Martins}, {Kudryashova}, {Kull}, {Bachchan},
  {Lacoste-Seris}, {Lanza}, {Lavigne}, {Le Poncin-Lafitte}, {Lebreton},
  {Lebzelter}, {Leccia}, {Leclerc}, {Lecoeur-Taibi}, {Lemaitre}, {Lenhardt},
  {Leroux}, {Liao}, {Licata}, {Lindstr{\o}m}, {Lister}, {Livanou}, {Lobel},
  {L{\"o}ffler}, {L{\'o}pez}, {Lopez-Lozano}, {Lorenz}, {Loureiro},
  {MacDonald}, {Magalh{\~a}es Fernandes}, {Managau}, {Mann}, {Mantelet},
  {Marchal}, {Marchant}, {Marconi}, {Marie}, {Marinoni}, {Marrese},
  {Marschalk{\'o}}, {Marshall}, {Mart{\'\i}n-Fleitas}, {Martino}, {Mary},
  {Matijevi{\v{c}}}, {Mazeh}, {McMillan}, {Messina}, {Mestre}, {Michalik},
  {Millar}, {Miranda}, {Molina}, {Molinaro}, {Molinaro}, {Moln{\'a}r},
  {Moniez}, {Montegriffo}, {Monteiro}, {Mor}, {Mora}, {Morbidelli}, {Morel},
  {Morgenthaler}, {Morley}, {Morris}, {Mulone}, {Muraveva}, {Musella},
  {Narbonne}, {Nelemans}, {Nicastro}, {Noval}, {Ord{\'e}novic},
  {Ordieres-Mer{\'e}}, {Osborne}, {Pagani}, {Pagano}, {Pailler}, {Palacin},
  {Palaversa}, {Parsons}, {Paulsen}, {Pecoraro}, {Pedrosa}, {Pentik{\"a}inen},
  {Pereira}, {Pichon}, {Piersimoni}, {Pineau}, {Plachy}, {Plum}, {Poujoulet},
  {Pr{\v{s}}a}, {Pulone}, {Ragaini}, {Rago}, {Rambaux}, {Ramos-Lerate},
  {Ranalli}, {Rauw}, {Read}, {Regibo}, {Renk}, {Reyl{\'e}}, {Ribeiro},
  {Rimoldini}, {Ripepi}, {Riva}, {Rixon}, {Roelens}, {Romero-G{\'o}mez},
  {Rowell}, {Royer}, {Rudolph}, {Ruiz-Dern}, {Sadowski}, {Sagrist{\`a}
  Sell{\'e}s}, {Sahlmann}, {Salgado}, {Salguero}, {Sarasso}, {Savietto},
  {Schnorhk}, {Schultheis}, {Sciacca}, {Segol}, {Segovia}, {Segransan},
  {Serpell}, {Shih}, {Smareglia}, {Smart}, {Smith}, {Solano}, {Solitro},
  {Sordo}, {Soria Nieto}, {Souchay}, {Spagna}, {Spoto}, {Stampa}, {Steele},
  {Steidelm{\"u}ller}, {Stephenson}, {Stoev}, {Suess}, {S{\"u}veges}, {Surdej},
  {Szabados}, {Szegedi-Elek}, {Tapiador}, {Taris}, {Tauran}, {Taylor},
  {Teixeira}, {Terrett}, {Tingley}, {Trager}, {Turon}, {Ulla}, {Utrilla},
  {Valentini}, {van Elteren}, {Van Hemelryck}, {van Leeuwen}, {Varadi},
  {Vecchiato}, {Veljanoski}, {Via}, {Vicente}, {Vogt}, {Voss}, {Votruba},
  {Voutsinas}, {Walmsley}, {Weiler}, {Weingrill}, {Werner}, {Wevers},
  {Whitehead}, {Wyrzykowski}, {Yoldas}, {{\v{Z}}erjal}, {Zucker}, {Zurbach},
  {Zwitter}, {Alecu}, {Allen}, {Allende Prieto}, {Amorim},
  {Anglada-Escud{\'e}}, {Arsenijevic}, {Azaz}, {Balm}, {Beck}, {Bernstein},
  {Bigot}, {Bijaoui}, {Blasco}, {Bonfigli}, {Bono}, {Boudreault}, {Bressan},
  {Brown}, {Brunet}, {Bunclark}, {Buonanno}, {Butkevich}, {Carret}, {Carrion},
  {Chemin}, {Ch{\'e}reau}, {Corcione}, {Darmigny}, {de Boer}, {de Teodoro}, {de
  Zeeuw}, {Delle Luche}, {Domingues}, {Dubath}, {Fodor}, {Fr{\'e}zouls},
  {Fries}, {Fustes}, {Fyfe}, {Gallardo}, {Gallegos}, {Gardiol}, {Gebran},
  {Gomboc}, {G{\'o}mez}, {Grux}, {Gueguen}, {Heyrovsky}, {Hoar}, {Iannicola},
  {Isasi Parache}, {Janotto}, {Joliet}, {Jonckheere}, {Keil}, {Kim},
  {Klagyivik}, {Klar}, {Knude}, {Kochukhov}, {Kolka}, {Kos}, {Kutka}, {Lainey},
  {LeBouquin}, {Liu}, {Loreggia}, {Makarov}, {Marseille}, {Martayan},
  {Martinez-Rubi}, {Massart}, {Meynadier}, {Mignot}, {Munari}, {Nguyen},
  {Nordlander}, {Ocvirk}, {O'Flaherty}, {Olias Sanz}, {Ortiz}, {Osorio},
  {Oszkiewicz}, {Ouzounis}, {Palmer}, {Park}, {Pasquato}, {Peltzer}, {Peralta},
  {P{\'e}turaud}, {Pieniluoma}, {Pigozzi}, {Poels}, {Prat}, {Prod'homme},
  {Raison}, {Rebordao}, {Risquez}, {Rocca-Volmerange}, {Rosen}, {Ruiz-Fuertes},
  {Russo}, {Sembay}, {Serraller Vizcaino}, {Short}, {Siebert}, {Silva},
  {Sinachopoulos}, {Slezak}, {Soffel}, {Sosnowska}, {Strai{\v{z}}ys}, {ter
  Linden}, {Terrell}, {Theil}, {Tiede}, {Troisi}, {Tsalmantza}, {Tur},
  {Vaccari}, {Vachier}, {Valles}, {Van Hamme}, {Veltz}, {Virtanen}, {Wallut},
  {Wichmann}, {Wilkinson}, {Ziaeepour}, \& {Zschocke}}]{Gaia16-Prusti}
{Gaia Collaboration}, {Prusti}, T., {de Bruijne}, J.~H.~J., {et~al.} 2016,
  \aap, 595, A1

\bibitem[{{Gaia Collaboration} {et~al.}(2017){Gaia Collaboration}, {van
  Leeuwen}, {Vallenari}, {Jordi}, {Lindegren}, {Bastian}, {Prusti}, {de
  Bruijne}, {Brown}, {Babusiaux}, {Bailer-Jones}, {Biermann}, {Evans}, {Eyer},
  {Jansen}, {Klioner}, {Lammers}, {Luri}, {Mignard}, {Panem}, {Pourbaix}, {Rand
  ich}, {Sartoretti}, {Siddiqui}, {Soubiran}, {Valette}, {Walton}, {Aerts},
  {Arenou}, {Cropper}, {Drimmel}, {H{\o}g}, {Katz}, {Lattanzi}, {O'Mullane},
  {Grebel}, {Holland }, {Huc}, {Passot}, {Perryman}, {Bramante}, {Cacciari},
  {Casta{\~n}eda}, {Chaoul}, {Cheek}, {De Angeli}, {Fabricius}, {Guerra},
  {Hern{\'a}ndez}, {Jean-Antoine-Piccolo}, {Masana}, {Messineo}, {Mowlavi},
  {Nienartowicz}, {Ord{\'o}{\~n}ez-Blanco}, {Panuzzo}, {Portell}, {Richards},
  {Riello}, {Seabroke}, {Tanga}, {Th{\'e}venin}, {Torra}, {Els},
  {Gracia-Abril}, {Comoretto}, {Garcia-Reinaldos}, {Lock}, {Mercier},
  {Altmann}, {Andrae}, {Astraatmadja}, {Bellas-Velidis}, {Benson}, {Berthier},
  {Blomme}, {Busso}, {Carry}, {Cellino}, {Clementini}, {Cowell}, {Creevey},
  {Cuypers}, {Davidson}, {De Ridder}, {de Torres}, {Delchambre}, {Dell'Oro},
  {Ducourant}, {Fr{\'e}mat}, {Garc{\'\i}a-Torres}, {Gosset}, {Halbwachs},
  {Hambly}, {Harrison}, {Hauser}, {Hestroffer}, {Hodgkin}, {Huckle}, {Hutton},
  {Jasniewicz}, {Jordan}, {Kontizas}, {Korn}, {Lanzafame}, {Manteiga},
  {Moitinho}, {Muinonen}, {Osinde}, {Pancino}, {Pauwels}, {Petit},
  {Recio-Blanco}, {Robin}, {Sarro}, {Siopis}, {Smith}, {Smith}, {Sozzetti},
  {Thuillot}, {van Reeven}, {Viala}, {Abbas}, {Abreu Aramburu}, {Accart},
  {Aguado}, {Allan}, {Allasia}, {Altavilla}, {{\'A}lvarez}, {Alves},
  {Anderson}, {Andrei}, {Anglada Varela}, {Antiche}, {Antoja}, {Ant{\'o}n},
  {Arcay}, {Bach}, {Baker}, {Balaguer-N{\'u}{\~n}ez}, {Barache}, {Barata},
  {Barbier}, {Barblan}, {Barrado y Navascu{\'e}s}, {Barros}, {Barstow},
  {Becciani}, {Bellazzini}, {Bello Garc{\'\i}a}, {Belokurov}, {Bendjoya},
  {Berihuete}, {Bianchi}, {Bienaym{\'e}}, {Billebaud}, {Blagorodnova},
  {Blanco-Cuaresma}, {Boch}, {Bombrun}, {Borrachero}, {Bouquillon}, {Bourda},
  {Bouy}, {Bragaglia}, {Breddels}, {Brouillet}, {Br{\"u}semeister},
  {Bucciarelli}, {Burgess}, {Burgon}, {Burlacu}, {Busonero}, {Buzzi}, {Caffau},
  {Cambras}, {Campbell}, {Cancelliere}, {Cantat-Gaudin}, {Carlucci},
  {Carrasco}, {Castellani}, {Charlot}, {Charnas}, {Chiavassa}, {Clotet},
  {Cocozza}, {Collins}, {Costigan}, {Crifo}, {Cross}, {Crosta}, {Crowley},
  {Dafonte}, {Damerdji}, {Dapergolas}, {David}, {David}, {De Cat}, {de Felice},
  {de Laverny}, {De Luise}, {De March}, {de Martino}, {de Souza}, {Debosscher},
  {del Pozo}, {Delbo}, {Delgado}, {Delgado}, {Di Matteo}, {Diakite},
  {Distefano}, {Dolding}, {Dos Anjos}, {Drazinos}, {Dur{\'a}n}, {Dzigan},
  {Edvardsson}, {Enke}, {Evans}, {Eynard Bontemps}, {Fabre}, {Fabrizio},
  {Faigler}, {Falc{\~a}o}, {Farr{\`a}s Casas}, {Federici}, {Fedorets},
  {Fern{\'a}ndez-Hern{\'a}ndez}, {Fernique}, {Fienga}, {Figueras}, {Filippi},
  {Findeisen}, {Fonti}, {Fouesneau}, {Fraile}, {Fraser}, {Fuchs}, {Gai},
  {Galleti}, {Galluccio}, {Garabato}, {Garc{\'\i}a-Sedano}, {Garofalo},
  {Garralda}, {Gavras}, {Gerssen}, {Geyer}, {Gilmore}, {Girona}, {Giuffrida},
  {Gomes}, {Gonz{\'a}lez-Marcos}, {Gonz{\'a}lez-N{\'u}{\~n}ez},
  {Gonz{\'a}lez-Vidal}, {Granvik}, {Guerrier}, {Guillout}, {Guiraud},
  {G{\'u}rpide}, {Guti{\'e}rrez-S{\'a}nchez}, {Guy}, {Haigron},
  {Hatzidimitriou}, {Haywood}, {Heiter}, {Helmi}, {Hobbs}, {Hofmann}, {Holl},
  {Holland }, {Hunt}, {Hypki}, {Icardi}, {Irwin}, {Jevardat de Fombelle},
  {Jofr{\'e}}, {Jonker}, {Jorissen}, {Julbe}, {Karampelas}, {Kochoska},
  {Kohley}, {Kolenberg}, {Kontizas}, {Koposov}, {Kordopatis}, {Koubsky},
  {Krone-Martins}, {Kudryashova}, {Kull}, {Bachchan}, {Lacoste-Seris}, {Lanza},
  {Lavigne}, {Le Poncin-Lafitte}, {Lebreton}, {Lebzelter}, {Leccia}, {Leclerc},
  {Lecoeur-Taibi}, {Lemaitre}, {Lenhardt}, {Leroux}, {Liao}, {Licata},
  {Lindstr{\o}m}, {Lister}, {Livanou}, {Lobel}, {L{\"o}ffler}, {L{\'o}pez},
  {Lorenz}, {MacDonald}, {Magalh{\~a}es Fernandes}, {Managau}, {Mann},
  {Mantelet}, {Marchal}, {Marchant}, {Marconi}, {Marinoni}, {Marrese},
  {Marschalk{\'o}}, {Marshall}, {Mart{\'\i}n-Fleitas}, {Martino}, {Mary},
  {Matijevi{\v{c}}}, {Mazeh}, {McMillan}, {Messina}, {Michalik}, {Millar},
  {Mirand a}, {Molina}, {Molinaro}, {Molinaro}, {Moln{\'a}r}, {Moniez},
  {Montegriffo}, {Mor}, {Mora}, {Morbidelli}, {Morel}, {Morgenthaler},
  {Morris}, {Mulone}, {Muraveva}, {Musella}, {Narbonne}, {Nelemans},
  {Nicastro}, {Noval}, {Ord{\'e}novic}, {Ordieres-Mer{\'e}}, {Osborne},
  {Pagani}, {Pagano}, {Pailler}, {Palacin}, {Palaversa}, {Parsons}, {Pecoraro},
  {Pedrosa}, {Pentik{\"a}inen}, {Pichon}, {Piersimoni}, {Pineau}, {Plachy},
  {Plum}, {Poujoulet}, {Pr{\v{s}}a}, {Pulone}, {Ragaini}, {Rago}, {Rambaux},
  {Ramos-Lerate}, {Ranalli}, {Rauw}, {Read}, {Regibo}, {Reyl{\'e}}, {Ribeiro},
  {Rimoldini}, {Ripepi}, {Riva}, {Rixon}, {Roelens}, {Romero-G{\'o}mez},
  {Rowell}, {Royer}, {Ruiz-Dern}, {Sadowski}, {Sagrist{\`a} Sell{\'e}s},
  {Sahlmann}, {Salgado}, {Salguero}, {Sarasso}, {Savietto}, {Schultheis},
  {Sciacca}, {Segol}, {Segovia}, {Segransan}, {Shih}, {Smareglia}, {Smart},
  {Solano}, {Solitro}, {Sordo}, {Soria Nieto}, {Souchay}, {Spagna}, {Spoto},
  {Stampa}, {Steele}, {Steidelm{\"u}ller}, {Stephenson}, {Stoev}, {Suess},
  {S{\"u}veges}, {Surdej}, {Szabados}, {Szegedi-Elek}, {Tapiador}, {Taris},
  {Tauran}, {Taylor}, {Teixeira}, {Terrett}, {Tingley}, {Trager}, {Turon},
  {Ulla}, {Utrilla}, {Valentini}, {van Elteren}, {Van Hemelryck}, {vanLeeuwen},
  {Varadi}, {Vecchiato}, {Veljanoski}, {Via}, {Vicente}, {Vogt}, {Voss},
  {Votruba}, {Voutsinas}, {Walmsley}, {Weiler}, {Weingrill}, {Wevers},
  {Wyrzykowski}, {Yoldas}, {{\v{Z}}erjal}, {Zucker}, {Zurbach}, {Zwitter},
  {Alecu}, {Allen}, {Allende Prieto}, {Amorim}, {Anglada-Escud{\'e}},
  {Arsenijevic}, {Azaz}, {Balm}, {Beck}, {Bernstein}, {Bigot}, {Bijaoui},
  {Blasco}, {Bonfigli}, {Bono}, {Boudreault}, {Bressan}, {Brown}, {Brunet},
  {Bunclark}, {Buonanno}, {Butkevich}, {Carret}, {Carrion}, {Chemin},
  {Ch{\'e}reau}, {Corcione}, {Darmigny}, {de Boer}, {de Teodoro}, {de Zeeuw},
  {Delle Luche}, {Domingues}, {Dubath}, {Fodor}, {Fr{\'e}zouls}, {Fries},
  {Fustes}, {Fyfe}, {Gallardo}, {Gallegos}, {Gardiol}, {Gebran}, {Gomboc},
  {G{\'o}mez}, {Grux}, {Gueguen}, {Heyrovsky}, {Hoar}, {Iannicola}, {Isasi
  Parache}, {Janotto}, {Joliet}, {Jonckheere}, {Keil}, {Kim}, {Klagyivik},
  {Klar}, {Knude}, {Kochukhov}, {Kolka}, {Kos}, {Kutka}, {Lainey}, {LeBouquin},
  {Liu}, {Loreggia}, {Makarov}, {Marseille}, {Martayan}, {Martinez-Rubi},
  {Massart}, {Meynadier}, {Mignot}, {Munari}, {Nguyen}, {Nordlander},
  {O'Flaherty}, {Ocvirk}, {Olias Sanz}, {Ortiz}, {Osorio}, {Oszkiewicz},
  {Ouzounis}, {Palmer}, {Park}, {Pasquato}, {Peltzer}, {Peralta},
  {P{\'e}turaud}, {Pieniluoma}, {Pigozzi}, {Poels}, {Prat}, {Prod'homme},
  {Raison}, {Rebordao}, {Risquez}, {Rocca-Volmerange}, {Rosen}, {Ruiz-Fuertes},
  {Russo}, {Sembay}, {Serraller Vizcaino}, {Short}, {Siebert}, {Silva},
  {Sinachopoulos}, {Slezak}, {Soffel}, {Sosnowska}, {Strai{\v{z}}ys}, {ter
  Linden}, {Terrell}, {Theil}, {Tiede}, {Troisi}, {Tsalmantza}, {Tur},
  {Vaccari}, {Vachier}, {Valles}, {Van Hamme}, {Veltz}, {Virtanen}, {Wallut},
  {Wichmann}, {Wilkinson}, {Ziaeepour}, \& {Zschocke}}]{Gaia17}
{Gaia Collaboration}, {van Leeuwen}, F., {Vallenari}, A., {et~al.} 2017, \aap,
  601, A19

\bibitem[{{Gaia Collaboration} {et~al.}(2018){Gaia Collaboration}, {Brown},
  {Vallenari}, {Prusti}, {de Bruijne}, {Babusiaux}, {Bailer-Jones}, {Biermann},
  {Evans}, {Eyer}, {Jansen}, {Jordi}, {Klioner}, {Lammers}, {Lindegren},
  {Luri}, {Mignard}, {Panem}, {Pourbaix}, {Randich}, {Sartoretti}, {Siddiqui},
  {Soubiran}, {van Leeuwen}, {Walton}, {Arenou}, {Bastian}, {Cropper},
  {Drimmel}, {Katz}, {Lattanzi}, {Bakker}, {Cacciari}, {Casta{\~n}eda},
  {Chaoul}, {Cheek}, {De Angeli}, {Fabricius}, {Guerra}, {Holl}, {Masana},
  {Messineo}, {Mowlavi}, {Nienartowicz}, {Panuzzo}, {Portell}, {Riello},
  {Seabroke}, {Tanga}, {Th{\'e}venin}, {Gracia-Abril}, {Comoretto},
  {Garcia-Reinaldos}, {Teyssier}, {Altmann}, {Andrae}, {Audard},
  {Bellas-Velidis}, {Benson}, {Berthier}, {Blomme}, {Burgess}, {Busso},
  {Carry}, {Cellino}, {Clementini}, {Clotet}, {Creevey}, {Davidson}, {De
  Ridder}, {Delchambre}, {Dell'Oro}, {Ducourant},
  {Fern{\'a}ndez-Hern{\'a}ndez}, {Fouesneau}, {Fr{\'e}mat}, {Galluccio},
  {Garc{\'\i}a-Torres}, {Gonz{\'a}lez-N{\'u}{\~n}ez}, {Gonz{\'a}lez-Vidal},
  {Gosset}, {Guy}, {Halbwachs}, {Hambly}, {Harrison}, {Hern{\'a}ndez},
  {Hestroffer}, {Hodgkin}, {Hutton}, {Jasniewicz}, {Jean-Antoine-Piccolo},
  {Jordan}, {Korn}, {Krone-Martins}, {Lanzafame}, {Lebzelter}, {L{\"o}ffler},
  {Manteiga}, {Marrese}, {Mart{\'\i}n-Fleitas}, {Moitinho}, {Mora}, {Muinonen},
  {Osinde}, {Pancino}, {Pauwels}, {Petit}, {Recio-Blanco}, {Richards},
  {Rimoldini}, {Robin}, {Sarro}, {Siopis}, {Smith}, {Sozzetti}, {S{\"u}veges},
  {Torra}, {van Reeven}, {Abbas}, {Abreu Aramburu}, {Accart}, {Aerts},
  {Altavilla}, {{\'A}lvarez}, {Alvarez}, {Alves}, {Anderson}, {Andrei},
  {Anglada Varela}, {Antiche}, {Antoja}, {Arcay}, {Astraatmadja}, {Bach},
  {Baker}, {Balaguer-N{\'u}{\~n}ez}, {Balm}, {Barache}, {Barata}, {Barbato},
  {Barblan}, {Barklem}, {Barrado}, {Barros}, {Barstow}, {Bartholom{\'e}
  Mu{\~n}oz}, {Bassilana}, {Becciani}, {Bellazzini}, {Berihuete}, {Bertone},
  {Bianchi}, {Bienaym{\'e}}, {Blanco-Cuaresma}, {Boch}, {Boeche}, {Bombrun},
  {Borrachero}, {Bossini}, {Bouquillon}, {Bourda}, {Bragaglia}, {Bramante},
  {Breddels}, {Bressan}, {Brouillet}, {Br{\"u}semeister}, {Brugaletta},
  {Bucciarelli}, {Burlacu}, {Busonero}, {Butkevich}, {Buzzi}, {Caffau},
  {Cancelliere}, {Cannizzaro}, {Cantat-Gaudin}, {Carballo}, {Carlucci},
  {Carrasco}, {Casamiquela}, {Castellani}, {Castro-Ginard}, {Charlot},
  {Chemin}, {Chiavassa}, {Cocozza}, {Costigan}, {Cowell}, {Crifo}, {Crosta},
  {Crowley}, {Cuypers}, {Dafonte}, {Damerdji}, {Dapergolas}, {David}, {David},
  {de Laverny}, {De Luise}, {De March}, {de Martino}, {de Souza}, {de Torres},
  {Debosscher}, {del Pozo}, {Delbo}, {Delgado}, {Delgado}, {Di Matteo},
  {Diakite}, {Diener}, {Distefano}, {Dolding}, {Drazinos}, {Dur{\'a}n},
  {Edvardsson}, {Enke}, {Eriksson}, {Esquej}, {Eynard Bontemps}, {Fabre},
  {Fabrizio}, {Faigler}, {Falc{\~a}o}, {Farr{\`a}s Casas}, {Federici},
  {Fedorets}, {Fernique}, {Figueras}, {Filippi}, {Findeisen}, {Fonti},
  {Fraile}, {Fraser}, {Fr{\'e}zouls}, {Gai}, {Galleti}, {Garabato},
  {Garc{\'\i}a-Sedano}, {Garofalo}, {Garralda}, {Gavel}, {Gavras}, {Gerssen},
  {Geyer}, {Giacobbe}, {Gilmore}, {Girona}, {Giuffrida}, {Glass}, {Gomes},
  {Granvik}, {Gueguen}, {Guerrier}, {Guiraud}, {Guti{\'e}rrez-S{\'a}nchez},
  {Haigron}, {Hatzidimitriou}, {Hauser}, {Haywood}, {Heiter}, {Helmi}, {Heu},
  {Hilger}, {Hobbs}, {Hofmann}, {Holland}, {Huckle}, {Hypki}, {Icardi},
  {Jan{\ss}en}, {Jevardat de Fombelle}, {Jonker}, {Juh{\'a}sz}, {Julbe},
  {Karampelas}, {Kewley}, {Klar}, {Kochoska}, {Kohley}, {Kolenberg},
  {Kontizas}, {Kontizas}, {Koposov}, {Kordopatis}, {Kostrzewa-Rutkowska},
  {Koubsky}, {Lambert}, {Lanza}, {Lasne}, {Lavigne}, {Le Fustec}, {Le
  Poncin-Lafitte}, {Lebreton}, {Leccia}, {Leclerc}, {Lecoeur-Taibi},
  {Lenhardt}, {Leroux}, {Liao}, {Licata}, {Lindstr{\o}m}, {Lister}, {Livanou},
  {Lobel}, {L{\'o}pez}, {Managau}, {Mann}, {Mantelet}, {Marchal}, {Marchant},
  {Marconi}, {Marinoni}, {Marschalk{\'o}}, {Marshall}, {Martino}, {Marton},
  {Mary}, {Massari}, {Matijevi{\v{c}}}, {Mazeh}, {McMillan}, {Messina},
  {Michalik}, {Millar}, {Molina}, {Molinaro}, {Moln{\'a}r}, {Montegriffo},
  {Mor}, {Morbidelli}, {Morel}, {Morris}, {Mulone}, {Muraveva}, {Musella},
  {Nelemans}, {Nicastro}, {Noval}, {O'Mullane}, {Ord{\'e}novic},
  {Ord{\'o}{\~n}ez-Blanco}, {Osborne}, {Pagani}, {Pagano}, {Pailler},
  {Palacin}, {Palaversa}, {Panahi}, {Pawlak}, {Piersimoni}, {Pineau}, {Plachy},
  {Plum}, {Poggio}, {Poujoulet}, {Pr{\v{s}}a}, {Pulone}, {Racero}, {Ragaini},
  {Rambaux}, {Ramos-Lerate}, {Regibo}, {Reyl{\'e}}, {Riclet}, {Ripepi}, {Riva},
  {Rivard}, {Rixon}, {Roegiers}, {Roelens}, {Romero-G{\'o}mez}, {Rowell},
  {Royer}, {Ruiz-Dern}, {Sadowski}, {Sagrist{\`a} Sell{\'e}s}, {Sahlmann},
  {Salgado}, {Salguero}, {Sanna}, {Santana-Ros}, {Sarasso}, {Savietto},
  {Schultheis}, {Sciacca}, {Segol}, {Segovia}, {S{\'e}gransan}, {Shih},
  {Siltala}, {Silva}, {Smart}, {Smith}, {Solano}, {Solitro}, {Sordo}, {Soria
  Nieto}, {Souchay}, {Spagna}, {Spoto}, {Stampa}, {Steele},
  {Steidelm{\"u}ller}, {Stephenson}, {Stoev}, {Suess}, {Surdej}, {Szabados},
  {Szegedi-Elek}, {Tapiador}, {Taris}, {Tauran}, {Taylor}, {Teixeira},
  {Terrett}, {Teyssand ier}, {Thuillot}, {Titarenko}, {Torra Clotet}, {Turon},
  {Ulla}, {Utrilla}, {Uzzi}, {Vaillant}, {Valentini}, {Valette}, {van Elteren},
  {Van Hemelryck}, {van Leeuwen}, {Vaschetto}, {Vecchiato}, {Veljanoski},
  {Viala}, {Vicente}, {Vogt}, {von Essen}, {Voss}, {Votruba}, {Voutsinas},
  {Walmsley}, {Weiler}, {Wertz}, {Wevers}, {Wyrzykowski}, {Yoldas},
  {{\v{Z}}erjal}, {Ziaeepour}, {Zorec}, {Zschocke}, {Zucker}, {Zurbach}, \&
  {Zwitter}}]{Gaia18-Brown}
{Gaia Collaboration}, {Brown}, A.~G.~A., {Vallenari}, A., {et~al.} 2018, \aap,
  616, A1

\bibitem[{{Hao} {et~al.}(2020){Hao}, {Xu}, {Wu}, {He}, \& {Bian}}]{Hao20}
{Hao}, C., {Xu}, Y., {Wu}, Z., {He}, Z., \& {Bian}, S. 2020, \pasp, 132, 034502

\bibitem[{Hunter(2007)}]{hunter07}
Hunter, J.~D. 2007, Computing in science \& engineering, 9, 90

\bibitem[{{Kharchenko} {et~al.}(2013){Kharchenko}, {Piskunov}, {Schilbach},
  {R{\"o}ser}, \& {Scholz}}]{Kharchenko13}
{Kharchenko}, N.~V., {Piskunov}, A.~E., {Schilbach}, E., {R{\"o}ser}, S., \&
  {Scholz}, R.~D. 2013, \aap, 558, A53

\bibitem[{{Lada} \& {Lada}(2003)}]{Lada03}
{Lada}, C.~J., \& {Lada}, E.~A. 2003, \araa, 41, 57

\bibitem[{{Lindegren} {et~al.}(2018){Lindegren}, {Hern{\'a}ndez}, {Bombrun},
  {Klioner}, {Bastian}, {Ramos-Lerate}, {de Torres}, {Steidelm{\"u}ller},
  {Stephenson}, {Hobbs}, {Lammers}, {Biermann}, {Geyer}, {Hilger}, {Michalik},
  {Stampa}, {McMillan}, {Casta{\~n}eda}, {Clotet}, {Comoretto}, {Davidson},
  {Fabricius}, {Gracia}, {Hambly}, {Hutton}, {Mora}, {Portell}, {van Leeuwen},
  {Abbas}, {Abreu}, {Altmann}, {Andrei}, {Anglada}, {Balaguer-N{\'u}{\~n}ez},
  {Barache}, {Becciani}, {Bertone}, {Bianchi}, {Bouquillon}, {Bourda},
  {Br{\"u}semeister}, {Bucciarelli}, {Busonero}, {Buzzi}, {Cancelliere},
  {Carlucci}, {Charlot}, {Cheek}, {Crosta}, {Crowley}, {de Bruijne}, {de
  Felice}, {Drimmel}, {Esquej}, {Fienga}, {Fraile}, {Gai}, {Garralda},
  {Gonz{\'a}lez-Vidal}, {Guerra}, {Hauser}, {Hofmann}, {Holl}, {Jordan},
  {Lattanzi}, {Lenhardt}, {Liao}, {Licata}, {Lister}, {L{\"o}ffler},
  {Marchant}, {Martin-Fleitas}, {Messineo}, {Mignard}, {Morbidelli}, {Poggio},
  {Riva}, {Rowell}, {Salguero}, {Sarasso}, {Sciacca}, {Siddiqui}, {Smart},
  {Spagna}, {Steele}, {Taris}, {Torra}, {van Elteren}, {van Reeven}, \&
  {Vecchiato}}]{Lindegren18}
{Lindegren}, L., {Hern{\'a}ndez}, J., {Bombrun}, A., {et~al.} 2018, \aap, 616,
  A2

\bibitem[{{Liu} \& {Pang}(2019)}]{Liu19}
{Liu}, L., \& {Pang}, X. 2019, \apjs, 245, 32

\bibitem[{{McKee} \& {Ostriker}(2007)}]{McKee07}
{McKee}, C.~F., \& {Ostriker}, E.~C. 2007, \araa, 45, 565

\bibitem[{McKinney {et~al.}(2010)}]{pandas}
McKinney, W., {et~al.} 2010, in Proceedings of the 9th Python in Science
  Conference, Vol. 445, Austin, TX, 51--56

\bibitem[{{O'Donnell}(1994)}]{Donnell94}
{O'Donnell}, J.~E. 1994, \apj, 422, 158

\bibitem[{Oliphant(2006)}]{numpy}
Oliphant, T.~E. 2006, A guide to NumPy, Vol.~1 (Trelgol Publishing USA)

\bibitem[{{Overbeek} {et~al.}(2017){Overbeek}, {Friel}, {Donati}, {Smiljanic},
  {Jacobson}, {Hatzidimitriou}, {Held}, {Magrini}, {Bragaglia}, {Randich},
  {Vallenari}, {Cantat-Gaudin}, {Tautvai{\v{s}}ien{\.{e}}},
  {Jim{\'e}nez-Esteban}, {Frasca}, {Geisler}, {Villanova}, {Tang}, {Mu{\~n}oz},
  {Marconi}, {Carraro}, {San Roman}, {Drazdauskas}, {{\v{Z}}enovien{\.{e}}},
  {Gilmore}, {Jeffries}, {Flaccomio}, {Pancino}, {Bayo}, {Costado}, {Damiani},
  {Jofr{\'e}}, {Monaco}, {Prisinzano}, {Sousa}, \& {Zaggia}}]{Overbeek17}
{Overbeek}, J.~C., {Friel}, E.~D., {Donati}, P., {et~al.} 2017, \aap, 598, A68

\bibitem[{{Pang} {et~al.}(2020){Pang}, {Li}, {Tang}, {Pasquato}, \&
  {Kouwenhoven}}]{Pang20}
{Pang}, X., {Li}, Y., {Tang}, S.-Y., {Pasquato}, M., \& {Kouwenhoven}, M.~B.~N.
  2020, \apjl, 900, L4

\bibitem[{Pedregosa {et~al.}(2011)Pedregosa, Varoquaux, Gramfort, Michel,
  Thirion, Grisel, Blondel, Prettenhofer, Weiss, Dubourg, Vanderplas, Passos,
  Cournapeau, Brucher, Perrot, \& Duchesnay}]{Pedregosa11}
Pedregosa, F., Varoquaux, G., Gramfort, A., {et~al.} 2011, Journal of Machine
  Learning Research, 12, 2825

\bibitem[{{Piskunov} {et~al.}(2006){Piskunov}, {Kharchenko}, {R{\"o}ser},
  {Schilbach}, \& {Scholz}}]{Piskunov06}
{Piskunov}, A.~E., {Kharchenko}, N.~V., {R{\"o}ser}, S., {Schilbach}, E., \&
  {Scholz}, R.~D. 2006, \aap, 445, 545

\bibitem[{Price-Whelan {et~al.}(2018)Price-Whelan, Sip{\H{o}}cz, G{\"u}nther,
  Lim, Crawford, Conseil, Shupe, Craig, Dencheva, Ginsburg, {et~al.}}]{astropy}
Price-Whelan, A.~M., Sip{\H{o}}cz, B., G{\"u}nther, H., {et~al.} 2018, The
  Astronomical Journal, 156, 123

\bibitem[{{Qin} {et~al.}(2020){Qin}, {Li}, {Chen}, \& {Zhong}}]{Qin20}
{Qin}, S.-m., {Li}, J., {Chen}, L., \& {Zhong}, J. 2020, arXiv e-prints,
  arXiv:2008.07164

\bibitem[{{S{\'a}nchez} {et~al.}(2010){S{\'a}nchez}, {Vicente}, \&
  {Alfaro}}]{Sanchez10}
{S{\'a}nchez}, N., {Vicente}, B., \& {Alfaro}, E.~J. 2010, \aap, 510, A78

\bibitem[{Sander {et~al.}(1998)Sander, Ester, Kriegel, \& Xu}]{Sander98}
Sander, J., Ester, M., Kriegel, H.-P., \& Xu, X. 1998, Data Min. Knowl.
  Discov., 2, 169

\bibitem[{{Sanders}(1971)}]{Sanders71}
{Sanders}, W.~L. 1971, \aap, 14, 226

\bibitem[{Schubert {et~al.}(2017)Schubert, Sander, Ester, Kriegel, \&
  Xu}]{Schubert17}
Schubert, E., Sander, J., Ester, M., Kriegel, H.-P., \& Xu, X. 2017, ACM Trans.
  Database Syst., 42, 19:1

\bibitem[{{Sim} {et~al.}(2019){Sim}, {Lee}, {Ann}, \& {Kim}}]{Sim19}
{Sim}, G., {Lee}, S.~H., {Ann}, H.~B., \& {Kim}, S. 2019, Journal of Korean
  Astronomical Society, 52, 145

\bibitem[{{Slovak}(1977)}]{Slovak77}
{Slovak}, M.~H. 1977, \aj, 82, 818

\bibitem[{{Uribe} \& {Brieva}(1994)}]{Uribe94}
{Uribe}, A., \& {Brieva}, E. 1994, \apss, 214, 171

\bibitem[{{Vereshchagin} \& {Chupina}(2016)}]{Vereshchagin16}
{Vereshchagin}, S.~V., \& {Chupina}, N.~V. 2016, Baltic Astronomy, 25, 432

\bibitem[{{Zhao} \& {He}(1990)}]{Zhao90}
{Zhao}, J.~L., \& {He}, Y.~P. 1990, \aap, 237, 54

\bibitem[{{Zu} \& {Zhao}(2003)}]{Zu03}
{Zu}, Z.~L., \& {Zhao}, J.~L. 2003, Progress in Astronomy, 21, 152

\end{thebibliography}

\appendix

\end{document}